\PassOptionsToPackage{unicode}{hyperref}
\PassOptionsToPackage{hyphens}{url}
\PassOptionsToPackage{dvipsnames,svgnames,x11names}{xcolor}
\documentclass[
  11pt,
  a4paper,
]{article}
\usepackage{xcolor}
\usepackage[left=2.5cm,right=2.5cm,top=2.5cm,bottom=3cm]{geometry}
\usepackage{amsmath,amssymb}
\usepackage{iftex}
\ifPDFTeX
  \usepackage[T1]{fontenc}
  \usepackage[utf8]{inputenc}
  \usepackage{textcomp} % provide euro and other symbols
\else % if luatex or xetex
  \ifXeTeX
    \usepackage{mathspec} % this also loads fontspec
  \else
    \usepackage{unicode-math} % this also loads fontspec
  \fi
  \defaultfontfeatures{Scale=MatchLowercase}
  \defaultfontfeatures[\rmfamily]{Ligatures=TeX,Scale=1}
\fi
\usepackage{lmodern}
\ifPDFTeX\else
\fi
\IfFileExists{upquote.sty}{\usepackage{upquote}}{}
\IfFileExists{microtype.sty}{% use microtype if available
  \usepackage[expansion=false]{microtype}
  \UseMicrotypeSet[protrusion]{basicmath} % disable protrusion for tt fonts
}{}
\makeatletter
\ifx\paragraph\undefined\else
  \let\oldparagraph\paragraph
  \renewcommand{\paragraph}{
    \@ifstar
      \xxxParagraphStar
      \xxxParagraphNoStar
  }
  \newcommand{\xxxParagraphStar}[1]{\oldparagraph*{#1}\mbox{}}
  \newcommand{\xxxParagraphNoStar}[1]{\oldparagraph{#1}\mbox{}}
\fi
\ifx\subparagraph\undefined\else
  \let\oldsubparagraph\subparagraph
  \renewcommand{\subparagraph}{
    \@ifstar
      \xxxSubParagraphStar
      \xxxSubParagraphNoStar
  }
  \newcommand{\xxxSubParagraphStar}[1]{\oldsubparagraph*{#1}\mbox{}}
  \newcommand{\xxxSubParagraphNoStar}[1]{\oldsubparagraph{#1}\mbox{}}
\fi
\makeatother

\usepackage{longtable,booktabs,array}
\usepackage{calc} % for calculating minipage widths
\usepackage{etoolbox}
\makeatletter
\patchcmd\longtable{\par}{\if@noskipsec\mbox{}\fi\par}{}{}
\makeatother
\IfFileExists{footnotehyper.sty}{\usepackage{footnotehyper}}{\usepackage{footnote}}
\makesavenoteenv{longtable}
\usepackage{graphicx}
\makeatletter
\newsavebox\pandoc@box
\newcommand*\pandocbounded[1]{% scales image to fit in text height/width
  \sbox\pandoc@box{#1}%
  \Gscale@div\@tempa{\textheight}{\dimexpr\ht\pandoc@box+\dp\pandoc@box\relax}%
  \Gscale@div\@tempb{\linewidth}{\wd\pandoc@box}%
  \ifdim\@tempb\p@<\@tempa\p@\let\@tempa\@tempb\fi% select the smaller of both
  \ifdim\@tempa\p@<\p@\scalebox{\@tempa}{\usebox\pandoc@box}%
  \else\usebox{\pandoc@box}%
  \fi%
}
\def\fps@figure{htbp}
\makeatother

\providecommand{\tightlist}{%
  \setlength{\itemsep}{0pt}\setlength{\parskip}{0pt}}

\usepackage[style=apa,backend=biber,sorting=nyt,doi=true,url=true,isbn=false,uniquename=false,uniquelist=false]{biblatex}
\usepackage{booktabs}
\usepackage{caption}
\usepackage{longtable}
\usepackage{colortbl}
\usepackage{array}
\usepackage{anyfontsize}
\usepackage{multirow}
\DeclareFieldFormat[article]{volume}{\textit{\apanum{#1}\nocorr}}
\DeclareFieldFormat[inproceedings,periodical]{volume}{\textit{\apanum{#1}\nocorr}}
\usepackage{setspace}
\newcommand{\N}{\mathrm{N}}
\DeclareMathOperator*{\argmax}{arg\,max}
\DeclareMathOperator*{\argmin}{arg\,min}
\usepackage{newunicodechar}
\usepackage{pifont}
\usepackage{textgreek}
\newunicodechar{✔}{\textnormal{\checkmark}}
\newunicodechar{δ}{\ensuremath{\delta}}
\newunicodechar{σ}{\ensuremath{\sigma}}
\AtBeginDocument{%
  \definecolor{kblu}{HTML}{003B75}%
  \definecolor{korg}{HTML}{804900}%
  \definecolor{ktur}{HTML}{004C59}%
  \definecolor{ktursat}{HTML}{00A6AA}% KAUST saturated turquoise
  \colorlet{kturlink}{ktursat!77!black}%
  \definecolor{quarto-callout-note-color}{HTML}{5284C4}%
  \definecolor{quarto-callout-note-color-frame}{HTML}{5284C4}%
  \hypersetup{citecolor=kturlink,linkcolor=kturlink,urlcolor=kturlink}%
}
\makeatletter
\AtBeginDocument{%
  \@ifundefined{c@lemma}{}{%
    \@ifundefined{c@corollary}{}{%
      \let\c@corollary\c@lemma
    }%
    \@ifundefined{c@proposition}{}{%
      \let\c@proposition\c@lemma
    }%
  }%
}
\makeatother
\makeatletter
\@ifpackageloaded{caption}{}{\usepackage{caption}}
\AtBeginDocument{%
\ifdefined\contentsname
  \renewcommand*\contentsname{Table of contents}
\else
  \newcommand\contentsname{Table of contents}
\fi
\ifdefined\listfigurename
  \renewcommand*\listfigurename{List of Figures}
\else
  \newcommand\listfigurename{List of Figures}
\fi
\ifdefined\listtablename
  \renewcommand*\listtablename{List of Tables}
\else
  \newcommand\listtablename{List of Tables}
\fi
\ifdefined\figurename
  \renewcommand*\figurename{Figure}
\else
  \newcommand\figurename{Figure}
\fi
\ifdefined\tablename
  \renewcommand*\tablename{Table}
\else
  \newcommand\tablename{Table}
\fi
}
\@ifpackageloaded{float}{}{\usepackage{float}}
\floatstyle{ruled}
\@ifundefined{c@chapter}{\newfloat{codelisting}{h}{lop}}{\newfloat{codelisting}{h}{lop}[chapter]}
\floatname{codelisting}{Listing}

\usepackage{amsthm}
\theoremstyle{plain}
\newtheorem{lemma}{Lemma}[section]
\theoremstyle{plain}
\newtheorem{proposition}{Proposition}[section]
\theoremstyle{plain}
\newtheorem{corollary}{Corollary}[section]
\theoremstyle{remark}
\AtBeginDocument{}

\makeatother
\makeatletter
\@ifpackageloaded{caption}{}{\usepackage{caption}}
\@ifpackageloaded{subcaption}{}{\usepackage{subcaption}}
\makeatother
\makeatletter
\@ifpackageloaded{sidenotes}{}{\usepackage{sidenotes}}
\@ifpackageloaded{marginnote}{}{\usepackage{marginnote}}
\makeatother
\usepackage{bookmark}
\IfFileExists{xurl.sty}{\usepackage{xurl}}{} % add URL line breaks if available
\hypersetup{
  pdftitle={Approximate Bayesian Inference for Structural Equation Models using Integrated Nested Laplace Approximations},
  pdfkeywords={Bayesian Structural Equation Model, Integrated Nested Laplace Approximation (INLA), Approximate Bayesian Inference, Variational Bayes, Skew-Normal Distribution},
  colorlinks=true,
  linkcolor={blue},
  filecolor={Maroon},
  citecolor={Blue},
  urlcolor={Blue},
  pdfcreator={LaTeX via pandoc}}

\title{Approximate Bayesian Inference for Structural Equation Models using Integrated Nested Laplace Approximations}
\usepackage{authblk}

\IfFileExists{orcidlink.sty}{\usepackage{orcidlink}}{\providecommand{\orcidlink}[1]{}}
\author[1,2]{Haziq Jamil~\orcidlink{0000-0003-3298-1010}\thanks{Correspondence concerning this article should be addressed to Haziq Jamil, Computer, Electrical and Mathematical Sciences and Engineering (CEMSE) Division, King Abdullah University of Science and Technology, Thuwal, 23955-6900, Kingdom of Saudi Arabia. E-mail: \href{mailto:haziq.jamil@kaust.edu.sa}{haziq.jamil@kaust.edu.sa}.}}
\author[1]{Håvard Rue~\orcidlink{0000-0002-0222-1881}}
\affil[1]{Computer, Electrical and Mathematical Sciences and Engineering (CEMSE) Division, King Abdullah University of Science and Technology}
\affil[2]{Mathematical Sciences, Faculty of Science, Universiti Brunei Darussalam}
\date{September 16, 2026}
\begin{document}
\maketitle
\begin{abstract}
Markov chain Monte Carlo (MCMC) methods remain the mainstay of Bayesian estimation of structural equation models (SEM), though they often incur a high computational cost. We present a bespoke approximate Bayesian approach to SEM, drawing on ideas from the integrated nested Laplace approximation (INLA; Rue et al., 2009, J. R. Stat. Soc., B: Stat. Methodol.) framework. After analytically integrating out the latent variables, we implement a simplified Laplace approximation that profiles the posterior using an efficient gradient-based volume correction, allowing for parametric skew-normal estimation of the marginals. A variational Bayes correction further refines the centre of the joint Gaussian approximation; marginal profiling implicitly recovers the same location correction to leading order. Essential quantities, including factor scores and model-fit indices, are obtained via a Gaussian copula sampling scheme. On a benchmark SEM, inference completes in seconds with close agreement to MCMC. A simulation study stresses the approximation along sample size, identification, boundary proximity, and model dimension. Accuracy remains high across the examined sample sizes and model dimensions, with degradation concentrated in weakly identified settings and in the lower tails of variances near the zero boundary.

\end{abstract}

\section{Introduction}\label{introduction}

Let \(\mathbf{y}_s \in \mathbb{R}^p\) denote the observed data vector for subject \(s = 1, \dots, n\), and define the stacked data vector \(\mathbf y = (\mathbf y_1^\top,\dots,\mathbf y_n^\top)^\top\).
Although not the dominant presentation in structural equation modelling (SEM) textbooks \autocite{kaplan2009structural,bollen1989structural,song2012basic,hoyle2023handbook}, it is quite natural, particularly in the Bayesian paradigm, to express SEMs as hierarchical latent Gaussian models \autocite[LGMs;][]{rue2009approximate}.
With \(\boldsymbol\eta = (\boldsymbol\eta_1^\top,\dots,\boldsymbol\eta_n^\top)^\top\), each latent vector \(\boldsymbol{\eta}_s \in \mathbb{R}^q\) being associated with subject \(s\), the data-generating process is described by the following nested system of probability distributions:
\begin{equation}\protect\phantomsection\label{eq-sem-lgm}{
\begin{aligned}
\mathbf y \mid \boldsymbol\eta, \boldsymbol\vartheta_1 &\sim \prod_{s=1}^n \phi_p(\mathbf y_s; \boldsymbol\nu + \boldsymbol\Lambda \boldsymbol\eta_s, \boldsymbol\Theta ) \\
\boldsymbol\eta \mid \boldsymbol\vartheta_2 &\sim \prod_{s=1}^n \phi_q\big((\mathbf I - \mathbf B)\boldsymbol\eta_s;  \boldsymbol\alpha, \boldsymbol\Psi \big) \\
\boldsymbol\vartheta := \{\boldsymbol\vartheta_1, \boldsymbol\vartheta_2\} &\sim \pi(\boldsymbol\vartheta),
\end{aligned}
}\end{equation}
where \(\phi_d(\cdot\, ; \boldsymbol\mu,\boldsymbol\Sigma)\) is the \(d\)-dimensional normal density with mean \(\boldsymbol\mu\) and covariance \(\boldsymbol\Sigma\), and \(\pi(\boldsymbol\vartheta)\) is a prior distribution over the model parameters \(\boldsymbol\vartheta \in \mathbb R^m\).

The SEM parameter vector \(\boldsymbol\vartheta\) collects the free entries of the vectors and matrices governing the first two levels of the hierarchy in (\ref{eq-sem-lgm}).
The first level specifies the \emph{measurement model}, parameterised conditionally on the latent variables \(\boldsymbol\eta\) by a \(p\)-vector of intercepts \(\boldsymbol\nu\), a \(p \times q\) factor loading matrix \(\boldsymbol\Lambda\) (typically sparse), and a \(p \times p\) error covariance matrix \(\boldsymbol\Theta\), their free entries forming \(\boldsymbol\vartheta_1\).
The second level defines the \emph{structural relations} among the latent variables, parameterised by the \(q \times q\) matrix of regression coefficients \(\mathbf B\), the \(q\)-vector of latent intercepts \(\boldsymbol\alpha\), and the \(q \times q\) covariance matrix of structural disturbances \(\boldsymbol\Psi\), their free entries forming \(\boldsymbol\vartheta_2\).
We assume measurement errors are independent of \(\boldsymbol\eta_s\), and take \(\mathbf B\) to be hollow (i.e., \(\operatorname{diag}(\mathbf B)=\mathbf 0_q\)).
Together with the usual scale-setting constraints (e.g., fixing one loading or factor variance per latent variable) and exclusion restrictions on \(\boldsymbol\Lambda\), \(\mathbf B\), and \(\boldsymbol\Psi\), these criteria yield a well-posed parameterisation that is likelihood-identified under standard regularity conditions \autocite{bollen2009two}.
Attention is restricted to recursive SEMs, where \(\mathbf B\) can be permuted to a strictly triangular form (equivalently, the directed graph is acyclic), making \((\mathbf I - \mathbf B)\) invertible, and hence the structural system algebraically well-defined \autocite{bentler1980linear}.

In the Bayesian framework, inference on the SEM targets mainly the posterior distribution
\begin{equation}\protect\phantomsection\label{eq-post-thetajoint}{
\pi(\boldsymbol{\vartheta} \mid \mathbf{y}) 
\propto \int 
\pi(\mathbf{y} \mid \boldsymbol{\eta} , \boldsymbol{\vartheta}) \, 
\pi(\boldsymbol{\eta} \mid \boldsymbol{\vartheta}) \, 
\pi(\boldsymbol{\vartheta}) \,
d \boldsymbol{\eta} .
}\end{equation}
Direct evaluation of this integral is generally intractable.
To circumvent high-dimensional integration, standard Markov chain Monte Carlo (MCMC) algorithms exploit the hierarchical structure of the LGM in (\ref{eq-sem-lgm}) by treating the latent variables \(\boldsymbol{\eta}\) as missing data to be sampled.
Conditioning on \(\boldsymbol{\eta}\) simplifies the likelihood, facilitating Gibbs or Metropolis-within-Gibbs sampling.
Consequently, this approach has been widely adopted in software such as Mplus \autocite{muthen2012bayesian} and early blavaan implementations in R \autocite{merkle2018blavaan} via JAGS \autocite{plummer2003jags}.
However, the method necessitates augmenting the state space with the vector \(\boldsymbol{\eta}\), whose dimension scales linearly with the sample size \(n\).
The resulting high dimensionality drives the substantial computational costs and poor mixing often observed in Bayesian SEM \autocite{ludtke2018more,hecht2019bayesian}, regardless of whether the person-specific scores \(\boldsymbol{\eta}\) are regarded as targets of inference or nuisance parameters \autocite{hecht2020integrating}.

The same hierarchical formulation in (\ref{eq-sem-lgm}) renders SEMs amenable to approximate Bayesian inference methods specifically designed for LGMs.
The most prominent of these is the integrated nested Laplace approximation \autocite[INLA;][]{rue2009approximate}.
INLA capitalises on the conditional independence properties of the latent field, formalising it as a Gaussian Markov random field \autocite[GMRF;][]{rue2005gaussian} often with sparse precision matrices, so key computations reduce to sparse linear solves and factorisations.
The core strategy of INLA for a generic LGM is to approximate posteriors of the kind in (\ref{eq-post-thetajoint}) by
\begin{equation}\protect\phantomsection\label{eq-sem-inla}{
\tilde\pi(\boldsymbol{\vartheta} \mid \mathbf{y}) 
\propto \frac{\pi(\mathbf y, \boldsymbol\eta, \boldsymbol\vartheta)}{\tilde\pi_G(\boldsymbol\eta \mid \mathbf y, \boldsymbol\vartheta)} \Bigg|_{\boldsymbol\eta = \boldsymbol\eta^*(\boldsymbol\vartheta)},
}\end{equation}
where \(\tilde\pi_G(\boldsymbol\eta \mid \mathbf y, \boldsymbol\vartheta)\) is a Gaussian approximation of the full conditional \(\pi(\boldsymbol{\eta} \mid \mathbf y, \boldsymbol{\vartheta})\), and \(\boldsymbol\eta^*(\boldsymbol\vartheta)\) is its mode.
The marginal components of (\ref{eq-sem-inla}) are further approximated by
\begin{equation}\protect\phantomsection\label{eq-marg-theta}{
\tilde\pi(\vartheta_j \mid \mathbf y) = \int \tilde\pi(\boldsymbol{\vartheta} \mid \mathbf{y}) \, d  \boldsymbol\vartheta_{-j}, \quad j=1,\dots,m,
}\end{equation}
and the posterior marginals for the latent variables by
\begin{equation}\protect\phantomsection\label{eq-marg-eta}{
\tilde\pi(\boldsymbol\eta_s \mid \mathbf y) = \int \tilde\pi(\boldsymbol\eta_s \mid \mathbf y, \boldsymbol{\vartheta}) \, \tilde\pi(\boldsymbol\vartheta \mid \mathbf y) \, d  \boldsymbol\vartheta,
}\end{equation}
in which \(\tilde\pi(\boldsymbol\eta_s \mid \mathbf y, \boldsymbol{\vartheta})\) is obtained through yet another Laplace-type approximation.
The efficiency of INLA stems from a carefully constructed sequence of nested Laplace approximations and low-dimensional numerical integration schemes that turn these integrals into accurate deterministic calculations.
For comprehensive reviews of the methodology, we refer readers to \textcite{martins2013bayesian}, \textcite{rue2017bayesian}, and \textcite{vanniekerk2023new}.

However, for the linear Gaussian SEMs under consideration, we contend that retaining \(\boldsymbol\eta\) as an explicit GMRF in the inference scheme is a suboptimal computational choice.
Unlike spatial or temporal models with dependence across a large latent field \autocite{krainski2018advanced}, SEMs typically have a latent field \(\boldsymbol\eta\) consisting of small, independent subject-level blocks, leaving little to gain from sparse-matrix algorithms.
Moreover, \(\boldsymbol\vartheta\) contains the substantive model parameters, forming a sizeable vector compared to the handful of hyperparameters typical of applications where INLA is most effective.
Consequently, the integrations in (\ref{eq-marg-theta}) and (\ref{eq-marg-eta}) require many posterior evaluations over \(\boldsymbol\vartheta\), each involving a factorisation of the latent-field precision matrix across all \(n\) subjects, making parameter exploration costly.

We propose to analytically integrate out the latent variables entirely and target the marginal likelihood \(\pi(\mathbf y \mid \boldsymbol\vartheta)\) directly.
Due to the Gaussian conjugacy of the first two levels in (\ref{eq-sem-lgm}), this marginal likelihood is also Gaussian, i.e., \(\pi(\mathbf y \mid \boldsymbol\vartheta) = \prod_{s=1}^n \phi_p(\mathbf y_s ; \boldsymbol\mu(\boldsymbol\vartheta), \boldsymbol\Sigma(\boldsymbol\vartheta))\), where

\vspace{-1em}

\begin{align}
\boldsymbol\mu(\boldsymbol\vartheta) &= \boldsymbol\nu + \boldsymbol\Lambda (\mathbf I - \mathbf B)^{-1} \boldsymbol\alpha \label{eq-reduced-form-a} \\
\boldsymbol\Sigma(\boldsymbol\vartheta) &= \boldsymbol\Lambda (\mathbf I - \mathbf B)^{-1} \boldsymbol\Psi (\mathbf I - \mathbf B)^{-\top} \boldsymbol\Lambda^\top + \boldsymbol\Theta. \label{eq-reduced-form-b}
\end{align}
\vspace{-1em}

Furthermore, the conditional posterior \(\pi(\boldsymbol\eta \mid \mathbf y, \boldsymbol\vartheta) = \prod_{s=1}^n \pi(\boldsymbol\eta_s \mid \mathbf y_s, \boldsymbol\vartheta)\) appearing in (\ref{eq-sem-inla}), whose factors appear in (\ref{eq-marg-eta}), is a product of exact Gaussian densities as well.
Each factor has a closed-form mean \(\mathbf m_s(\boldsymbol\vartheta)\) and covariance \(\mathbf V(\boldsymbol\vartheta)\) given by
\begin{equation}\protect\phantomsection\label{eq-cond-post-eta}{
\begin{aligned}
\mathbf m_s(\boldsymbol\vartheta) &= 
\mathbf V\big(\boldsymbol\Phi^{-1}(\mathbf I - \mathbf B)^{-1}\boldsymbol\alpha +\boldsymbol\Lambda^\top\boldsymbol\Theta^{-1}(\mathbf y_s-\boldsymbol\nu)\big) \\
\mathbf V(\boldsymbol\vartheta) &= \big(\boldsymbol\Phi^{-1}+\boldsymbol\Lambda^\top\boldsymbol\Theta^{-1}\boldsymbol\Lambda\big)^{-1} \\
\boldsymbol\Phi &:= (\mathbf I-\mathbf B)^{-1}\boldsymbol\Psi(\mathbf I-\mathbf B)^{-\top}.
\end{aligned}
}\end{equation}
This ``collapsed'' perspective effectively reduces the inference problem to the \(m\)-dimensional space of \(\boldsymbol\vartheta\), bypassing the need for numerical treatment of the high-dimensional latent field.
The latent variables can easily be recovered, either via empirical Bayes plug-in estimates or post-hoc sampling.

This change in perspective parallels the evolution of the blavaan R package.
\textcite{merkle2021efficient} noted that their initial data-augmentation implementation was often slow for standard models.
Shifting to a marginal likelihood approach for standard Gaussian SEMs using Stan \autocite{carpenter2017stan} improved sampling efficiency and convergence stability, despite prevailing literature advocating latent-variable sampling for its simplicity \autocite[e.g.,][]{lee2007structural}.

The primary contribution of this article is a framework for fast and accurate approximation of marginal posterior densities in normal-theory SEMs (Section~\ref{sec-methodology}), adapting key ideas from INLA to the SEM setting.
At the core are complementary joint and marginal posterior approximations.
The joint Gaussian approximation obtained by Laplace's method is refined through a variational Bayes correction to its centre.
Marginal densities follow from simplified Laplace profiling, with parametric skew-normal fits capturing asymmetry.
Finally, an efficient Gaussian copula sampling scheme is employed to propagate uncertainty to derived quantities of interest, including factor scores and model-fit indices.

The scope of the framework is worth stating at the outset.
Everything that follows rests on the latent variables integrating out analytically, leaving a marginal likelihood that is multivariate Gaussian as in {\eqref{eq-reduced-form-a} and \eqref{eq-reduced-form-b}}.
Any specification that preserves this property is accommodated, and this single condition admits a wide class of SEMs, including multigroup models, growth parameterisations, and fixed-value or equality constraints.
Under ignorable missingness, missing responses are handled through the observed-data likelihood, with each case contributing a Gaussian density over its observed coordinates.
Exogenous covariates may be included in the joint Gaussian model or, when fully observed, conditioned on as fixed; in the latter case, they may be continuous or discrete.
These specifications are standard extensions of the covariance structure model, all within the scope of normal-theory SEM \autocite{bollen1989structural}.
Non-Gaussian response models, including binary, ordinal, and count models, fall outside the present scope and are discussed later.

The remainder of the paper is organised as follows.
Section~\ref{sec-preliminaries} reviews the mathematical background on Laplace's method and the skew-normal distribution.
Section~\ref{sec-methodology} introduces the proposed methodology.
Section~\ref{sec-validation} benchmarks the approximation against MCMC, isolates the contribution of each layer of the marginal construction through an ablation, and examines operating conditions and calibration through simulation.
Section~\ref{sec-discussion} discusses implications and future directions, and Section~\ref{sec-conclusion} concludes.

\section{Preliminaries}\label{sec-preliminaries}

This section reviews the technical foundations underlying the proposed methodology.

\subsection{Laplace's Method}\label{laplaces-method}

Let \(x\in \mathbb R^d\) and \(y\in\mathbb R^n\) be vectors.
In this subsection, vectors and matrices are shown in non-bold typeface for legibility.
Laplace's method \autocite[Section 3.3]{barndorff1989asymptotic} provides accurate approximations to integrals of the form
\[
I = \int_{\mathbb R^d} f(y, x) \, d x, \quad f(y,x) > 0
\]
by exploiting the fact that, when \(f(y, \cdot)\) is sufficiently concentrated, the integral is dominated by a neighbourhood around its maximiser.
Assume that for each fixed \(y\), \(\log f(y,x)\) has a unique mode \(x^* := x^*(y)\) and that the negative Hessian at the mode,
\[
H = - \nabla^2_x \log f(y,x) \Big|_{x = x^*},
\]
is positive definite (\(H \succ 0\)).
A second-order Taylor expansion of \(\log f(y,x)\) about \(x^*\) yields the Gaussian (quadratic) approximation, and integrating the resulting kernel with respect to \(x\) gives
\[
I \approx (2\pi)^{d/2} |H|^{-1/2} f\big(y,x^*\big),
\]
or equivalently,
\[
I \approx \frac{f(y,x = x^*)}{ \tilde\pi_G(x =  x^* \mid y)},
\]
where \(\tilde\pi_G(x \mid y)\) is the Gaussian density with mean \(x^*\) and precision \(H(y)\).

Suppose that the integrand admits the large-sample representation
\[
I_n = \int g(y,x) \exp\{n L(y,x) \} \, d x, \qquad L(y,x) = \frac{1}{n}\log f_n(y,x),
\]
and that the standard regularity conditions of \textcite{tierney1986accurate} hold, namely a unique interior mode, local smoothness, and exponential tail decay.
Then, Laplace's method yields an approximation \(\tilde I_n\) that is asymptotically accurate with relative error \(O(n^{-1})\), i.e., \(I_n = \tilde I_n\{1 + O(n^{-1}) \}\) \autocite{rue2017bayesian}, and hence \(\log I_n = \log \tilde I_n + O(n^{-1})\).
The accuracy of the Gaussian approximation is underpinned by the Bernstein--von Mises theorem \autocite[see, e.g.,][]{vandervaart1998asymptotic}:
for fixed dimension, the posterior concentrates at rate \(n^{-1/2}\) and converges to a Gaussian in total variation, so a second-order expansion of the log-posterior around its mode becomes increasingly accurate.

\subsection{Skew-Normal Distribution}\label{skew-normal-distribution}

The skew-normal (SN) distribution \autocite{azzalini1985class} extends the normal distribution by introducing a shape parameter that governs asymmetry, while retaining many of the tractable properties of the Gaussian family.

A random variable \(X\) follows a skew-normal distribution, written \(X \sim \mathrm{SN}(\xi, \omega, \alpha)\), if its probability density function is
\[
f_{\mathrm{SN}}(x; \xi, \omega, \alpha) = \frac{2}{\omega}\,\phi\!\left(\frac{x - \xi}{\omega}\right)\,\Phi\!\left(\alpha\,\frac{x - \xi}{\omega}\right), \quad x \in \mathbb{R},
\]
where \(\phi(\cdot)\) and \(\Phi(\cdot)\) denote the standard normal density and distribution function, respectively.
The three parameters have the following roles:

\begin{itemize}
\tightlist
\item
  \(\xi \in \mathbb{R}\) is a location parameter;
\item
  \(\omega > 0\) is a scale parameter; and
\item
  \(\alpha \in \mathbb{R}\) is a shape (skewness) parameter.
\end{itemize}

When \(\alpha = 0\) the density reduces to \(\N(\xi, \omega^2)\); positive (negative) values of \(\alpha\) induce right (left) skewness.
The corresponding cumulative distribution function (CDF) does not admit a simple closed-form expression but can be written as
\[
F_{\mathrm{SN}}(x; \xi, \omega, \alpha) = \Phi\!\left(\frac{x - \xi}{\omega}\right) - 2\,T\!\left(\frac{x - \xi}{\omega},\, \alpha\right),
\]
where \(T(h, a) = (2\pi)^{-1}\int_0^a \exp \bigl\{-\tfrac{1}{2}h^2(1+t^2)\bigr\}/(1+t^2)\,dt\) is Owen's \(T\)-function \autocite{owen1956tables}.

Define \(\delta = \alpha / \sqrt{1 + \alpha^2}\).
The mean, variance, and skewness of \(X\) are
\begin{equation}\protect\phantomsection\label{eq-sn-moments}{
\operatorname{E}(X) = \xi + \omega \delta \sqrt{\frac{2}{\pi}}, 
\qquad
\operatorname{Var}(X) = \omega^2\left(1 - \frac{2\delta^2}{\pi}\right),
\qquad
\operatorname{Skew}(X) = \frac{4 - \pi}{2}\,\frac{\bigl(\delta\sqrt{2/\pi}\bigr)^3}{\bigl(1 - 2\delta^2/\pi\bigr)^{3/2}}.
}\end{equation}
The coefficient of skewness of the skew-normal family is bounded in the interval \((-0.9953, 0.9953)\) \autocite{azzalini1985class}, which limits the degree of asymmetry it can represent.
In practice, this range is more than adequate for the posterior marginals encountered in typical Bayesian applications.

\section{Posterior Approximation}\label{sec-methodology}

\begin{figure}[p]

\centering{

\includegraphics[width=0.99\linewidth,height=\textheight,keepaspectratio]{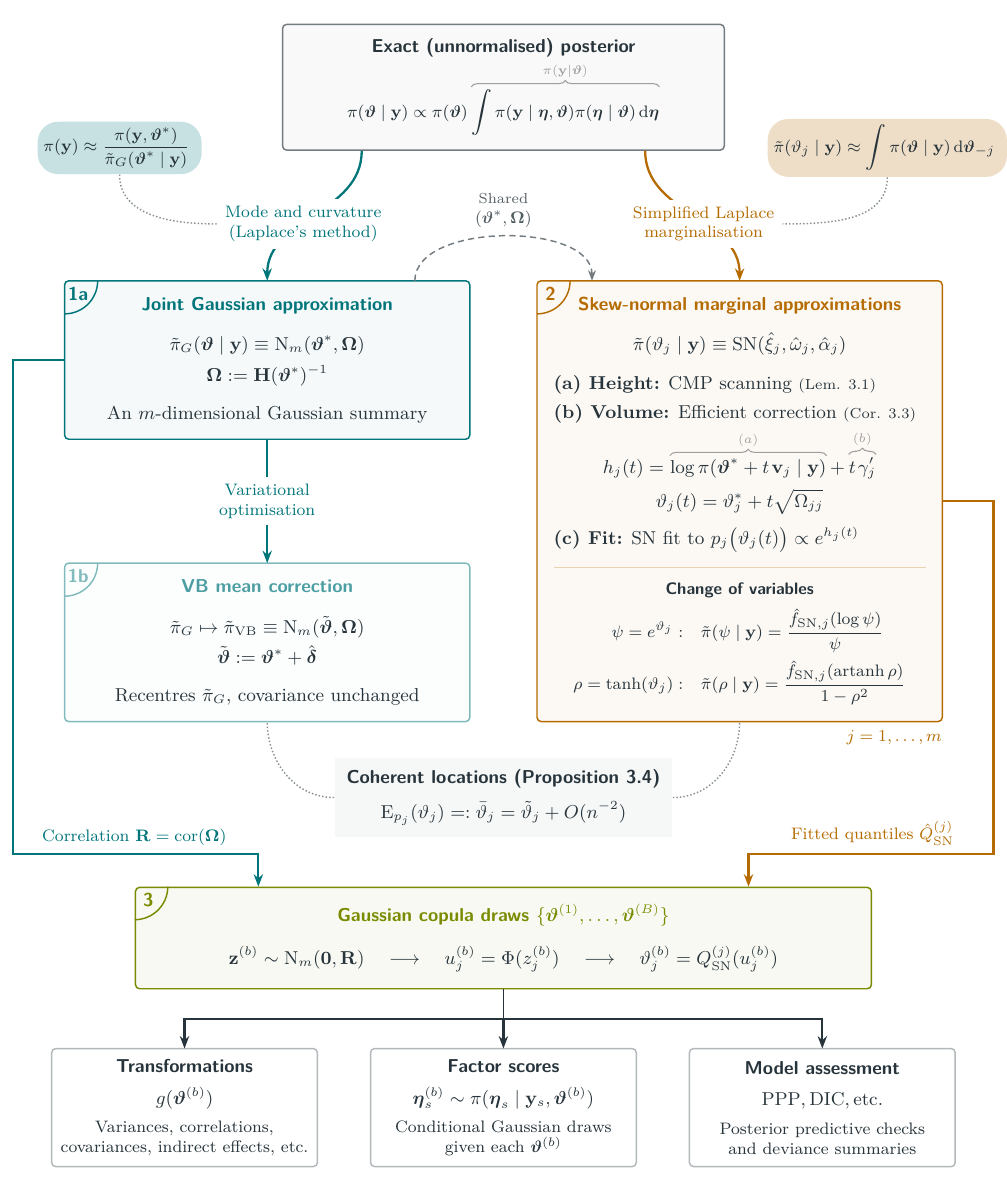}

}

\caption{\label{fig-process}Overview of the posterior approximation scheme. Laplace's method is applied to the exact posterior in Equation~\ref{eq-post-thetajoint} in two ways: to the joint posterior, which yields the Gaussian approximation (1a) and its variational mean correction (1b), and to the integral defining each marginal, which yields the skew-normal fits to the marginal profiles (2). The Gaussian copula (3) combines the fitted marginal quantiles with the correlation structure of \(\boldsymbol\Omega\) into joint draws, from which transformations, factor scores, and model-fit quantities are computed.}

\end{figure}

We now describe a procedure for accurate, approximate Bayesian inference for normal-theory SEMs, consisting of three main components.
First, we construct a joint Gaussian approximation to the posterior using Laplace's method, refined by a variational Bayes (VB) correction of its mean.
Second, we perform deterministic marginalisation of the exact posterior that circumvents high-dimensional numerical integration.
This involves marginal profiling via axis scanning, volume correction of the profile densities, and skew-normal curve fitting to obtain smooth, normalised marginal densities.
Finally, we implement a Gaussian copula sampling scheme to estimate derived quantities of interest.
Figure~\ref{fig-process} summarises the scheme and the flow of quantities between its components.

\subsection{Joint Gaussian Approximation With Variational Mean Correction}\label{sec-vb}

In Bayesian analysis, Laplace's method is used to approximate intractable posterior normalising constants and posterior marginals \autocite{tierney1986accurate}.
Write the posterior density for the SEM in (\ref{eq-post-thetajoint}) as
\[
\pi(\boldsymbol\vartheta \mid \mathbf y) = \frac{\pi(\mathbf y \mid \boldsymbol\vartheta)  \, \pi(\boldsymbol\vartheta)}{\pi(\mathbf y)}, \qquad \pi(\mathbf y) = \int \pi(\mathbf y , \boldsymbol\vartheta) \, d \boldsymbol\vartheta,
\]
using the Gaussian likelihood for \(\boldsymbol\vartheta \in \mathbb R^m\) with mean and covariance as specified in
{\eqref{eq-reduced-form-a} and \eqref{eq-reduced-form-b}}
respectively.

We then see that evaluating \(\pi(\mathbf y)\) involves an integral of the Laplace form.
An approximation \(\tilde\pi_G(\boldsymbol\vartheta\mid \mathbf y)\) to the posterior is thus \emph{locally} Gaussian centred at the \emph{maximum a posteriori} estimate \(\boldsymbol\vartheta^* = \argmax_{\boldsymbol\vartheta} \log \pi(\mathbf y,\boldsymbol\vartheta)\) having precision
\begin{equation}\protect\phantomsection\label{eq-joint-lap-prec}{
\mathbf H := \mathbf H(\boldsymbol\vartheta^*) = - \nabla^2_{\boldsymbol\vartheta} \log \pi(\mathbf y,\boldsymbol\vartheta) \Big|_{\boldsymbol\vartheta = \boldsymbol\vartheta^*} \in \mathbb R^{m\times m}, \quad \mathbf H \succ \mathbf 0.
}\end{equation}
That is, \(\tilde\pi_G(\boldsymbol\vartheta \mid \mathbf y) = \phi_m(\boldsymbol\vartheta; \boldsymbol\vartheta^*, \boldsymbol\Omega)\) with covariance \(\boldsymbol\Omega := \mathbf H^{-1}\).

In implementation, parameters subject to positivity or boundedness constraints (e.g., variances, correlations) are mapped to \(\mathbb{R}\) via standard differentiable transformations (log, Fisher), and all mode-finding, Hessian evaluation, and later on, marginal profiling, are carried out in this unconstrained parameter space.
In principle, the approximation scheme requires an unconstrained parameterisation under which every \(\boldsymbol\vartheta \in \mathbb R^m\) defines an admissible SEM; the present implementation handles inadmissible configurations numerically.
These transformations also improve the local quadratic approximation, since a Laplace approximation is most accurate when the log-posterior is close to quadratic.
A variance parameter, for instance, is often heavily right-skewed on its natural scale but nearly bell-shaped on the log scale.

The joint Laplace approximation can fail to capture the typical set when the posterior mode is not a representative centre of mass, leading to point summaries that sit away from the bulk of the posterior probability.
As a simple, low-cost remedy, we adopt the VB mean-shift correction of \textcite{vanniekerk2024lowrank}, which keeps the covariance fixed at its Laplace value \(\boldsymbol\Omega\) and chooses a location shift \(\boldsymbol\delta\in\mathbb R^m\) that minimises the Kullback--Leibler divergence (KLD) to the true posterior,
\begin{equation}\protect\phantomsection\label{eq-kld}{
\hat{\boldsymbol\delta}
=
\argmin_{\boldsymbol\delta} \,
D_{\mathrm{KL}}
\big(
\phi_m(\cdot ; \boldsymbol\vartheta^*+\boldsymbol\delta,\boldsymbol\Omega)
\ \Vert \
\pi(\cdot \mid \mathbf y)
\big).
}\end{equation}
Restricting the approximating density \(q\) to Gaussians with covariance \(\boldsymbol\Omega\) is a pragmatic choice, since the Laplace step has already accounted for the curvature and leaves only the centre to be corrected.
What is in principle a search across a space of probability densities thus condenses into a straightforward optimisation over \(\boldsymbol\delta\), bypassing the cost of re-estimating an \(m\times m\) covariance matrix.
Minimising the KLD provides, in the sense of \textcite{zellner1988optimal}, the optimal information-processing rule for summarising the posterior distribution \(\pi(\cdot \mid \mathbf y)\) by a tractable proxy \(q\).

The optimisation in (\ref{eq-kld}) is equivalent to maximising an expected log-posterior under the shifted Gaussian.
Indeed,
\[
D_{\mathrm{KL}}\left(
q_{\boldsymbol\delta} \, \Vert \, \pi(\cdot\mid \mathbf y)
\right)
=
\operatorname{E}_{q_{\boldsymbol\delta}} \big\{
\log q_{\boldsymbol\delta} (\boldsymbol\vartheta) \big\}
- \operatorname{E}_{q_{\boldsymbol\delta}} \big\{\log \pi(\boldsymbol\vartheta\mid \mathbf y)\big\},
\]
and since \(q_{\boldsymbol\delta}\) has covariance \(\boldsymbol\Omega\) fixed, its entropy (hence \(\operatorname{E}_{q_{\boldsymbol\delta}}(\log q_{\boldsymbol\delta})\)) is constant in \(\boldsymbol\delta\).
Therefore \(\hat{\boldsymbol\delta}\) can be obtained by maximising the expected log-posterior,
\begin{equation}\protect\phantomsection\label{eq-vb-objective}{
\hat{\boldsymbol\delta}
=
\argmax_{\boldsymbol\delta} \,
\operatorname{E}_{\boldsymbol\vartheta\sim \N_m(\boldsymbol\vartheta^*+\boldsymbol\delta,\boldsymbol\Omega)}
\big\{
\log \pi(\boldsymbol\vartheta\mid \mathbf y)
\big\}.
}\end{equation}
Since \(\boldsymbol\vartheta\) enters the likelihood nonlinearly through \(\boldsymbol\mu(\boldsymbol\vartheta)\) and \(\boldsymbol\Sigma(\boldsymbol\vartheta)\), the expectation in (\ref{eq-vb-objective}) has no closed form and requires numerical evaluation.
This can be approximated reliably using a randomised quasi-Monte Carlo rule \autocite[RQMC;][]{lecuyer2018randomized}, e.g., via scrambled Sobol points mapped elementwise to \(\N(0,1)\) by a probit transform, yielding a low-discrepancy sample-average objective.
We may also carry out the maximisation (\ref{eq-vb-objective}) in a whitened parameterisation, which typically improves numerical conditioning.

The VB mean shift thus defines the corrected joint Gaussian approximation \(\N_m(\tilde{\boldsymbol\vartheta},\boldsymbol\Omega)\), where \(\tilde{\boldsymbol\vartheta} := \boldsymbol\vartheta^* + \hat{\boldsymbol\delta}\), which replaces the mode-centred approximation in downstream steps involving it.
Calculation of the \emph{model evidence}, or marginal data density, is a case in point.
Approximating the identity \(\pi(\mathbf y) = \pi(\mathbf y, \boldsymbol\vartheta) / \pi(\boldsymbol\vartheta \mid \mathbf y)\) at the centre \(\boldsymbol\vartheta = \tilde{\boldsymbol\vartheta}\), with the corrected Gaussian in place of the exact posterior, gives, in log form,
\[
\log \tilde\pi(\mathbf y) = \frac{m}{2}\log (2\pi) - \frac{1}{2}\log |\mathbf H| + \log \pi(\mathbf y \mid \tilde{\boldsymbol\vartheta}) + \log \pi(\tilde{\boldsymbol\vartheta}),
\]
which, since \(\hat{\boldsymbol\delta}=O(n^{-1})\) (see the proof of Proposition~\ref{prp-vb-coherence}), satisfies \(\log \pi(\mathbf y) = \log \tilde\pi(\mathbf y) + O(n^{-1})\) under the conditions of Section~\ref{sec-preliminaries}.
This reduces to the usual Laplace approximation of the evidence when \(\hat{\boldsymbol\delta} = \mathbf 0\).

\subsection{Skew-Normal Laplace Profiling of Posterior Marginals}\label{sec-marginals}

If \(\pi(\boldsymbol\vartheta \mid \mathbf y)\) were Gaussian, the joint Laplace approximation would be exact, and the univariate marginals would follow immediately from standard properties of the multivariate normal distribution.
In practice, SEM posteriors often exhibit skewness and other departures from normality, particularly for variance and scale parameters \autocite{muthen2012bayesian}.
Curvature at the mode alone cannot capture global asymmetry, with the discrepancy propagating to the marginal distributions.

We instead evaluate the Laplace-profiled log-marginal \(\log \tilde{\pi}(\vartheta_j \mid \mathbf y)\) on a grid for each \(j\), and fit a skew-normal density to these evaluations.
Throughout, the expansions are taken about the mode \(\boldsymbol\vartheta^*\) (where the log-posterior is stationary) rather than at the corrected centre \(\tilde{\boldsymbol\vartheta}\), and the location adjustment this appears to forgo is in fact recovered to leading order---see Proposition~\ref{prp-vb-coherence} below.
For each component \(j=1,\dots,m\), the exact posterior marginal is
\[
\pi(\vartheta_j \mid \mathbf{y}) = \int \pi(\vartheta_j, \boldsymbol\vartheta_{-j} \mid \mathbf{y}) \, d \boldsymbol\vartheta_{-j},
\]
where \(\boldsymbol\vartheta_{-j}\) denotes the vector of parameters \(\boldsymbol\vartheta\) with its \(j\)th element removed.
Applying Laplace's method yet again to this integral gives the second-order approximation
\[
\log \tilde{\pi}(\vartheta_j \mid \mathbf{y}) 
= \log \pi(\vartheta_j, \boldsymbol\vartheta_{-j}^* \mid \mathbf{y}) - \frac{1}{2} \log |\mathbf H_{-j}(\vartheta_j)| + \frac{m-1}{2}\log(2\pi).
\]
Direct evaluation of this approximation is, however, costly.
For each grid point \(\vartheta_j\) one must (a) solve an \((m-1)\)-dimensional optimisation problem over the slice \(\{\vartheta_j\} \times \mathbb R^{m-1}\) to obtain \(\boldsymbol\vartheta_{-j}^*(\vartheta_j)\), and (b) compute \(\log | \mathbf H_{-j}(\vartheta_j)|\), where \(\boldsymbol\vartheta_{-j}^* := \boldsymbol\vartheta_{-j}^*(\vartheta_j)\) is the conditional maximiser and \(\mathbf H_{-j}(\vartheta_{j})\) is the corresponding \((m-1) \times (m-1)\) negative Hessian matrix.
The following subsections describe a more efficient strategy that avoids repeated optimisations and Hessian evaluations while retaining salient marginal features, including skewness.

\subsubsection{Conditional Mean Path}\label{conditional-mean-path}

To profile the \(j\)th component efficiently, we replace the slice-wise optimisations by a deterministic trajectory through \(\mathbb R^m\) that tracks the ridge of high posterior density as \(\vartheta_j\) varies.
This trajectory is the \emph{conditional mean path (CMP)}, defined as the locus of points where the remaining \((m-1)\) parameters \(\boldsymbol\vartheta_{-j}\) are set to their conditional expectations given \(\vartheta_j\):
\[
\mathcal{C}_j
=
\big\{
(x, \boldsymbol\vartheta_{-j}) \in \mathbb{R}^m
\;\big|\;
x \in \mathbb R, \,
\boldsymbol\vartheta_{-j}
=
\operatorname{E}_{\tilde\pi_G} (\boldsymbol\vartheta_{-j}\mid \vartheta_j = x)
\big\}.
\]
Here, expectations are taken with respect to the Gaussian approximation \(\tilde\pi_G(\boldsymbol\vartheta\mid \mathbf y)\) from the joint Laplace step, and are available in closed form.
The rationale is that, for a multivariate normal target, marginalisation in \(\boldsymbol\vartheta_{-j}\) is equivalent (up to proportionality) to evaluation along the CMP, as formalised in the following lemma \autocite[which also appears as Lemma 1 in Section 3.2.2 of][]{martins2013bayesian}.

\begin{lemma}[Marginalisation via the CMP]\protect\hypertarget{lem-mvn-marginal-slice}{}\label{lem-mvn-marginal-slice}

Let \(\boldsymbol\vartheta\sim\N_m(\boldsymbol\vartheta^*,\boldsymbol\Omega)\) with \(\boldsymbol\Omega \succ \mathbf 0\).
Then the marginal density of the \(j\)th component, \(j=1,\dots,m\), is proportional to the joint density evaluated along its CMP:
\[
\pi(\vartheta_j=x)
\propto
\pi\big(\vartheta_j=x, \, \boldsymbol\vartheta_{-j} = \operatorname{E}(\boldsymbol\vartheta_{-j} \mid \vartheta_j=x) \big).
\]

\end{lemma}

To see this, note that the multivariate normal conditional expectation is affine,
\[
\operatorname{E}(\boldsymbol\vartheta \mid \vartheta_j = x)
=
\boldsymbol\vartheta^* + (x-\vartheta_j^*)\,\frac{\boldsymbol\Omega_{\cdot j}}{\Omega_{jj}},
\]
where \(\boldsymbol\Omega_{\cdot j}\) is the \(j\)th column of \(\boldsymbol\Omega\). Substituting into the joint quadratic form reduces it to \((x-\vartheta_j^*)^2/\Omega_{jj}\), recovering the marginal kernel.
The lemma provides an operational insight:
for a Gaussian density, the \emph{exact} marginal of \(\vartheta_j\) can be recovered by slicing the joint density along the regression line of \(\boldsymbol\vartheta\) on \(\vartheta_j\).
For a non-Gaussian posterior, scanning along this linear trajectory serves as a first-order approximation to a potentially curved ridge.

We therefore define our scan direction as follows.
Recall that \(\boldsymbol\Omega = \mathbf H^{-1}\) is the covariance matrix of the joint Laplace approximation, with \(\mathbf H\) as in (\ref{eq-joint-lap-prec}).
To facilitate profiling, define the scan trajectory and the \(j\)th grid of \(K\) points along it,
\begin{equation}\protect\phantomsection\label{eq-theta-grid}{
\boldsymbol\vartheta(t) = \boldsymbol\vartheta^* + t\,\mathbf v_j, \qquad
\mathcal G_j = \big\{ \boldsymbol\vartheta(t_k) \big\}_{k=1}^K, \quad t_k \in [-4,4], \qquad
\mathbf v_j = \frac{\boldsymbol\Omega_{\cdot j}}{\sqrt{\Omega_{jj}}}.
}\end{equation}
Here, the direction vector \(\mathbf v_j\) is normalised so that \(t\) acts as a \(Z\)-score for \(\vartheta_j\); a unit step \(t=1\) induces a displacement of one standard deviation (SD), \(\sqrt{\Omega_{jj}}\), in the \(j\)th coordinate, while the remaining \((m-1)\) components adjust according to their linear dependence on \(\vartheta_j\) via \(\boldsymbol\Omega_{\cdot j}\).
The interval \([-4,4]\) thus corresponds to \(\vartheta_j^*\pm4\sqrt{\Omega_{jj}}\), a range that would hold 99.994\% of the mass of a Gaussian with that scale.

\subsubsection{Efficient Volume Correction}\label{sec-efficient-volume}

Evaluating \(\log \pi(\boldsymbol\vartheta\mid \mathbf y)\) along \(\mathcal G_j\) replaces the Gaussian quadratic decay implied by the joint Laplace approximation with a proxy for the slice-wise peak log-density (log-``height'') as \(\vartheta_j\) varies.
Accurate marginalisation, however, also requires the Laplace volume correction, which enters on the log scale as \(-\tfrac12\log|\mathbf H_{-j}(\vartheta_j)|\), and captures changes in the conditional spread of \(\boldsymbol\vartheta_{-j}\) orthogonal to the scan direction.
We now show how to update this volume correction efficiently along the grid without full Hessian re-evaluations.

Consider again the scan trajectory \(\boldsymbol\vartheta(t)=\boldsymbol\vartheta^*+t\,\mathbf v_j\) used to profile the \(j\)th posterior marginal on the grid \(\mathcal G_j\).
Write \(\mathbf H_{-j}(t)\) for the \((m-1)\times(m-1)\) submatrix of \(\mathbf H(t) := \mathbf H(\boldsymbol\vartheta(t))\) with the \(j\)th row and column removed, and define \(\gamma_j(t) := -\frac12 \log\big|\mathbf H_{-j}(t)\big|\).
A first-order expansion yields \(\gamma_j(t) \approx \gamma_j(0)+t\,\gamma_j'(0)\), so that along the trajectory the volume correction can be extrapolated linearly from the slope \(\gamma_j'(0)\).
This linear term tilts the profile, shifting mass towards the side on which the conditional spread of \(\boldsymbol\vartheta_{-j}\) widens.
The constant \(\gamma_j(0)\) contributes only an additive shift to the log-marginal along \(\mathcal G_j\), and is handled by the intercept (normalising) term in the subsequent skew-normal fit.

As a remark, higher-order expansions of \(\gamma_j(t)\) are possible, but we focus on the first-order update here; implications are discussed in Section~\ref{sec-discussion}.
The following lemma shows that the slope \(\gamma_j'(0)\) admits the form of a projected trace of the whitened Hessian perturbation at the mode.

\begin{lemma}[Volume-Slope Decomposition]\protect\hypertarget{lem-eff-vol}{}\label{lem-eff-vol}

Let \(\mathbf z=\mathbf L^{-1}(\boldsymbol\vartheta-\boldsymbol\vartheta^*)\) with \(\mathbf H(\boldsymbol\vartheta^*)^{-1}=\mathbf L\mathbf L^\top\), and write \(\mathbf J(t)=\mathbf L^\top \mathbf H(t)\,\mathbf L\) for the negative Hessian in whitened coordinates along the scan trajectory.
Let \(\mathbf w_j = \mathbf L^{-1}\mathbf v_j\) denote the image of the scan direction in the whitened frame.
Then \(\mathbf J := \mathbf J(0)=\mathbf I_m\), \(\|\mathbf w_j\| = 1\), and
\begin{equation}\protect\phantomsection\label{eq-vol-slope}{
\gamma_j' = -\frac12 \operatorname{tr}\left(\mathbf J'\right) + \frac12\,\mathbf w_j^\top \mathbf J'\,\mathbf w_j,
}\end{equation}
where \(\mathbf J' := \mathbf J'(0) = \frac{d}{dt}\mathbf J(t)\big|_{t=0}\) and, likewise, \(\gamma_j' := \gamma_j'(0)\).

\end{lemma}

Writing \(\mathbf P_j^\perp = \mathbf I_m - \mathbf w_j\mathbf w_j^\top\) for the projector onto the orthogonal complement of \(\mathbf w_j\), the volume slope becomes
\[
\gamma_j' = -\frac12\operatorname{tr}\left(\mathbf P_j^\perp\,\mathbf J'\right).
\]
The geometric content is now apparent: the volume slope measures the rate of curvature change in the whitened frame, restricted to directions transverse to the scan.
Since \(\mathbf J = \mathbf I_m\), the whitening renders all baseline curvatures unity, so the formula reduces entirely to traces of the first-order perturbation \(\mathbf J'\).
And this result is \emph{independent} of the choice of the whitener \(\mathbf L\).
The next corollary shows how these quantities can be evaluated efficiently using only gradient information.

\begin{corollary}[Gradient Representation of the Volume Slope]\protect\hypertarget{cor-eff-vol}{}\label{cor-eff-vol}

Let \(\mathbf g(\boldsymbol\vartheta) = -\nabla\log \pi(\boldsymbol\vartheta\mid \mathbf y)\), so that \(\mathbf H(\boldsymbol\vartheta) = \nabla\mathbf g(\boldsymbol\vartheta)\) and \(\mathbf u^\top\mathbf H(\boldsymbol\vartheta)\,\mathbf u = \mathbf u^\top\frac{d}{d\mathbf u}\mathbf g(\boldsymbol\vartheta)\) for any fixed direction \(\mathbf u\).
Then the volume slope (\ref{eq-vol-slope}) admits the gradient representation
\begin{equation}\protect\phantomsection\label{eq-vol-slope-gradient}{
\gamma_j'
=
-\frac12 \sum_{k=1}^m
\frac{d}{dt}\!\left[
\mathbf L_{\cdot k}^{\top}\,\frac{d}{d\mathbf L_{\cdot k}}\mathbf g
\right]_{t=0}
+
\frac12\,
\frac{d}{dt}\!\left[
\mathbf v_j^{\top}\,\frac{d}{d\mathbf v_j}\mathbf g
\right]_{t=0},
}\end{equation}
where \(\mathbf L_{\cdot k}\) is the \(k\)th column of \(\mathbf L\).
The \(m\) summands resolve the trace in (\ref{eq-vol-slope}), and the final term is the Schur correction.

\end{corollary}

Evaluating each part in (\ref{eq-vol-slope-gradient}) involves two finite-difference layers, both anchored at the mode where \(\mathbf H(\boldsymbol\vartheta^*) = (\mathbf L\mathbf L^\top)^{-1}\):
an \emph{inner} perturbation by step size \(\varepsilon_{\mathrm{in}}\) along each of the \(m+1\) directions \(\mathbf u \in \{\mathbf L_{\cdot 1},\dots,\mathbf L_{\cdot m},\mathbf v_j\}\), approximating the bracketed terms \(\frac{d}{d\mathbf u}\mathbf g\); and an \emph{outer} perturbation by step size \(\varepsilon_{\mathrm{out}}\) along the scan direction \(\mathbf v_j\), approximating the derivative in \(t\).
Both reduce to evaluations of \(\mathbf g(\boldsymbol\vartheta)\) alone, without ever forming or factoring \(\mathbf H(\boldsymbol\vartheta)\).
Note that the score of the SEM log-likelihood is a standard closed-form result \autocite{neudecker1991linear,vonoertzen2014efficient}, so provided the log-priors are twice differentiable, \(\mathbf g\) is exact and the differencing introduces error only one derivative order above the log-posterior.
Since the whitening renders all \(m+1\) baseline curvatures unity at the mode, a forward-difference scheme is particularly economical.
The outer layer needs only a single shifted point \(\boldsymbol\vartheta^+=\boldsymbol\vartheta^*+\varepsilon_{\mathrm{out}}\,\mathbf v_j\), differenced against the known baseline of 1 at the mode.
At \(\boldsymbol\vartheta^+\), a single base gradient \(\mathbf g(\boldsymbol\vartheta^+)\) is shared across all \(m+1\) inner perturbations, each differenced against one additional evaluation \(\mathbf g(\boldsymbol\vartheta^+ + \varepsilon_{\mathrm{in}}\mathbf u)\).
The total cost is therefore just \(m+2\) gradient evaluations per scan direction \(j\).
Both layers can be independently upgraded to central differences for improved accuracy.
Upgrading the inner layer alone costs \(2(m+1)\) evaluations, as the shared base at \(\boldsymbol\vartheta^+\) is replaced by a symmetric pair around each direction.

\subsubsection{Skew-Normal Curve Fitting}\label{skew-normal-curve-fitting}

The preceding steps provide approximate marginal log-density ordinates on a grid.
We convert these ordinates into a smooth, normalised approximation by fitting a skew-normal density, which provides a unimodal parametric family capable of capturing moderate skewness.

Fix \(j\in\{1,\dots,m\}\).
Recall the grid \(\mathcal G_j = \{\boldsymbol\vartheta(t_k)\}_{k=1}^K\) from (\ref{eq-theta-grid}), ordered along the scan direction \(\mathbf v_j\).
Combining the height evaluation with the linearised volume correction gives the unnormalised, volume-corrected log-marginal along the scan, and evaluating it at the grid points gives the ordinates for the fit:
\begin{equation}\protect\phantomsection\label{eq-corrected-ordinates}{
h_j(t) := \log\pi\big(\boldsymbol\vartheta(t) \mid \mathbf y\big) + t\,\gamma_j',
\qquad
h_{kj} := h_j(t_k), \quad k = 1,\dots,K,
}\end{equation}
where \(\gamma_j'\) is the slope of the volume correction derived in the previous subsection.

From the pairs \(\{(\vartheta_j(t_k),h_{kj})\}_{k=1}^K\), we approximate the marginal density of \(\vartheta_j\) by a skew-normal distribution \(\mathrm{SN}(\xi_j,\omega_j,\alpha_j)\).
Since the values \(h_{kj}\) are log-density ordinates defined only up to an additive constant, we include a free intercept \(c_j\) and fit by weighted least squares,
\[
(\hat\xi_j,\hat\omega_j,\hat\alpha_j,\hat c_j)
=
\argmin_{\xi,\omega,\alpha,c} \,
\sum_{k=1}^K
w_{kj}\big[
h_{kj}-\big\{\log f_{\mathrm{SN}}(\vartheta_j(t_k);\xi,\omega,\alpha)+c\big\}
\big]^2 .
\]
A convenient default is to weight points by their implied mass, \(w_{kj} \propto \exp(h_{kj})\), which prioritises fidelity near the mode; alternative weightings may be used to emphasise tail behaviour.
With the normalised approximation \(\tilde\pi(\vartheta_j \mid \mathbf y) = f_{\mathrm{SN}}(\vartheta_j;\hat\xi_j,\hat\omega_j,\hat\alpha_j)\), posterior summaries (means, SDs, and quantiles) are computed using standard formulae \autocite[e.g.,][]{azzalini1985class} or by direct numerical inversion of the fitted CDF.

It is worth pausing to ask why marginalisation was anchored at the mode \(\boldsymbol\vartheta^*\) and not at the VB-corrected centre \(\tilde{\boldsymbol\vartheta}\) that Section~\ref{sec-vb} worked to obtain.
The preceding subsections suggest part of the answer.
Profiling and volume correction each move the marginal mean away from the mode, through the skew of the height term along the CMP and the pull of the volume tilt on the marginal mass.
Thus, the marginal can be displaced from \(\vartheta_j^*\) even without an explicit correction.
The question, then, is by how much.
As it turns out, the VB shift coincides with the \(m\)-dimensional counterpart of this mode-to-mean displacement to leading order.
The following proposition formalises this relationship.

\begin{proposition}[Equivalence of Location Corrections]\protect\hypertarget{prp-vb-coherence}{}\label{prp-vb-coherence}

Let \(\bar\vartheta_j\) be the marginal mean under \(p_j(t)\propto\exp\{h_j(t)\}\) on the scale of \(\vartheta_j\), with \(h_j\) as in (\ref{eq-corrected-ordinates}).
Under the regularity conditions in Section~\ref{sec-preliminaries}, as \(n\to\infty\) with \(m\) fixed,
\[
\bar\vartheta_j = \vartheta_j^* + \hat\delta_j + O(n^{-2}), \quad j=1,\dots,m,
\]
where \(\hat\delta_j\) is the \(j\)th component of the VB mean correction (\ref{eq-vb-objective}).

\end{proposition}

The proof, given in the appendix, connects both corrections to the classical \(O(n^{-1})\) mode-to-mean displacement of the posterior, determined by the curvature and third derivatives of the log-posterior at the mode \autocite{lindley1980approximate,tierney1986accurate}.
The VB shift estimates this displacement jointly via an \(m\)-dimensional optimisation, whereas the mode-anchored marginal scheme recovers it componentwise through the combined effect of profile asymmetry and the volume tilt.
Thus, the joint and marginal approximations are automatically coherent in location.
Operationally, the scans remain anchored at the mode and no separate VB translation need be introduced in the marginal construction; the VB shift serves only to recentre the joint Gaussian.

\subsection{Gaussian Copula Sampling}\label{sec-copula}

The preceding steps deliver closed-form approximations for the posterior marginals \(\pi(\vartheta_j\mid \mathbf y)\), \(j=1,\dots,m\), via fitted skew-normal distributions.
For inference that depends only on a single-component monotonic transformation of them (e.g., \(\vartheta_j \mapsto \exp(\vartheta_j)\) for variances), these marginals suffice.
Quantiles follow by transformation of the fitted CDF, and one-dimensional expectations can be evaluated directly (e.g., by quadrature).

For inference involving transformations of multiple parameters simultaneously, we require an approximation to the joint dependence structure.
One could sample directly from the corrected joint Gaussian \(\N_m(\tilde{\boldsymbol\vartheta},\boldsymbol\Omega)\), but this would discard the marginal accuracy gained in the preceding steps, reverting each component to a Gaussian marginal.
Since our methodology targets marginal accuracy while retaining only a local Gaussian approximation of dependence at the mode, we instead reconstruct joint posterior samples using a Gaussian copula \autocite{nelsen2006introduction} with the fitted skew-normal marginals.
Let \(\boldsymbol\Omega = \mathbf H(\boldsymbol\vartheta^*)^{-1}\) denote the covariance matrix from the joint Laplace approximation, and define the associated correlation matrix
\[
\mathbf R = \mathbf D^{-1/2}\,\boldsymbol\Omega\,\mathbf D^{-1/2},
\qquad \mathbf D=\operatorname{diag}(\boldsymbol\Omega).
\]
Let \(F_{\mathrm{SN}}^{(j)}\) denote the fitted skew-normal CDF for the \(j\)th marginal, and \(Q_{\mathrm{SN}}^{(j)}(u) = \inf \{x \mid F_{\mathrm{SN}}^{(j)}(x) \geq u \}\) the corresponding quantile function.
For \(b=1,\dots,B\), a joint draw \(\boldsymbol\vartheta^{(b)}\) is obtained by:

\begin{enumerate}
\def\labelenumi{\arabic{enumi}.}
\tightlist
\item
  \textbf{Copula draw:} Sample \(\mathbf z^{(b)}\sim \N_m(\mathbf 0,\mathbf R)\);
\item
  \textbf{Probability integral transform:} Set \(u_j^{(b)}=\Phi(z_j^{(b)})\), \(j=1,\dots,m\);
\item
  \textbf{Marginal inversion:} Set \(\vartheta_j^{(b)}=Q_{\mathrm{SN}}^{(j)}(u_j^{(b)})\), \(j=1,\dots,m\).
\end{enumerate}

\noindent Thus, only the correlation structure is inherited from the joint Gaussian, while location and scale are determined by the fitted marginals.

By construction, the resulting draws \(\{\boldsymbol\vartheta^{(b)}\}_{b=1}^B\) preserve the fitted marginals, so sample-based quantities remain consistent with the reported marginal summaries.
The joint draws can also be transformed to approximate the induced posterior distribution of a derived quantity \(g(\boldsymbol\vartheta)\), including functions of several parameters such as an implied covariance or indirect effect.
For a scalar quantity \(g:\mathbb R^m\rightarrow\mathbb R\), a skew-normal distribution may be fitted to the transformed draws \(\{g(\boldsymbol\vartheta^{(b)})\}_{b=1}^B\), when appropriate, to provide a compact parametric approximation.
One limitation is that the reconstructed joint posterior is restricted to Gaussian-copula dependence and therefore cannot represent asymmetric copula dependence or nonzero asymptotic tail dependence.

The marginal inversion in Step 3 preserves the Gaussian copula and its rank-based dependence measures \autocite{nelsen2006introduction}, but reduces the magnitude of Pearson correlations and never inflates or reverses them \autocite{lancaster1957properties}.
The size of the reduction for any \((j,k)\) pair is a deterministic function of the correlation \(\rho := R_{jk}\) and the two skew-normal shape parameters, and can be evaluated exactly.
Let \(\zeta_l(z) = \{Q_{\mathrm{SN}}^{(l)}(\Phi(z)) - \mu_l\}/\sigma_l\), \(l \in \{j, k\}\), denote the inverted marginal standardised by its mean and standard deviation from (\ref{eq-sn-moments}), and let \((Z_j, Z_k)\) be standard bivariate normal with correlation \(\rho\).
The Pearson correlation of the transformed pair is then \(\tilde\rho = \operatorname{E}\{\zeta_j(Z_j)\,\zeta_k(Z_k)\}\), which two-dimensional Gauss--Hermite quadrature evaluates to machine precision.
Over most of the family the reduction \(\rho - \tilde\rho\) is small.
For instance, at the skewness bound of \(0.9953\) for both marginals (\(|\alpha_j|, |\alpha_k| \to \infty\), the half-normal limit) and \(\rho = 0.5\) it is \(0.019\) when the skews share the sign of \(\rho\) and \(0.054\) when they do not.
The reduction exceeds \(0.1\) only for the latter configuration at \(|\rho| \geq 0.8\).
In such cases, a NORmal-To-Anything adjustment \autocite[NORTA;][]{cario1997modeling} can recover the Laplace correlations by calibrating those supplied to Step 1.
Outside these cases, the reductions are too small to warrant the adjustment, so we do not pursue it here.

\section{Accuracy, Efficiency, and Calibration}\label{sec-validation}

This section presents the validation study.
First, we describe the benchmark model, prior specification, and computing environment.
We then compare the proposed approximation against a high-accuracy MCMC reference and use an ablation study to isolate the contribution of each layer of the marginal construction.
Finally, we examine the method's operating conditions through a repeated-sampling simulation and close with a simulation-based calibration check.

\subsection{Benchmark Model, Priors, and Computing Environment}\label{benchmark-model-priors-and-computing-environment}

\begin{figure}[htbp]

\centering{

\includegraphics[width=0.97\linewidth,height=\textheight,keepaspectratio]{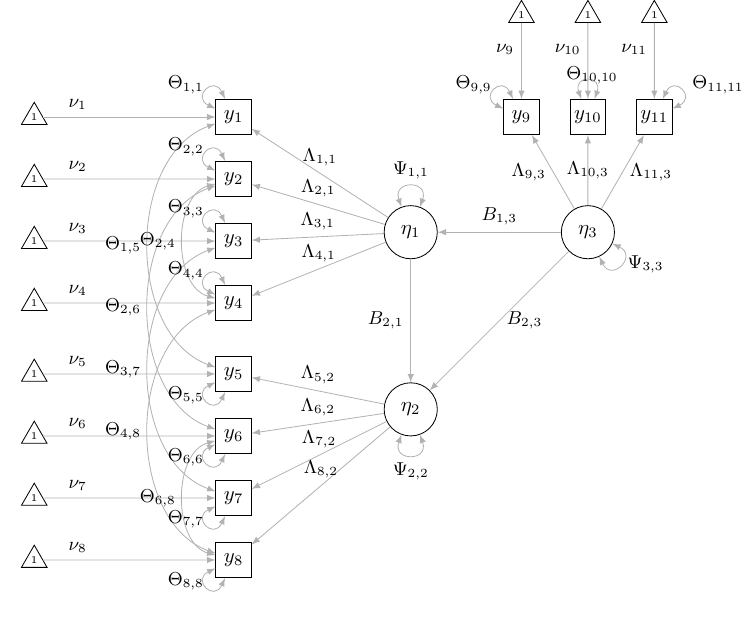}

}

\caption{\label{fig-poldem}Path diagram for the Political Democracy SEM \autocite{bollen1989structural}. Circles denote latent variables, squares denote observed indicators, and triangles denote the constant 1. Single-headed arrows represent intercepts (\(\nu_i\)), factor loadings (\(\Lambda_{i,j}\)), and structural regression coefficients (\(B_{j,j'}\)). Double-headed arrows represent residual variances (\(\Theta_{i,i}\)), residual covariances (\(\Theta_{i,j}\), \(i \neq j\)), and latent variances (\(\Psi_{j,j}\)). The diagram depicts the model for a single observation; the same structure is assumed independently for each \(s = 1, \ldots, n\).}

\end{figure}

We use the Political Democracy model of \textcite{bollen1989structural} as a benchmark, owing to its long-standing role as a canonical SEM with both measurement and structural components.
The model comprises three latent variables measured by multiple observed indicators, coupled with structural regressions linking them (see Figure~\ref{fig-poldem}).
Its ubiquity in the SEM literature and the availability of widely reported estimates make it a standard reference point for assessing approximate Bayesian inference.
In total, \(m=42\) parameters are estimated in this SEM, including the means of the observations.
It is worth noting that the Political Democracy model, with its modest sample size (\(n = 75\)), non-trivial structural component, and six residual covariances, constitutes a challenging benchmark for any approximate method.
Good performance on this benchmark would therefore be an encouraging indication of how the method fares on the class of SEMs most commonly encountered in practice.

Priors are specified to be weakly regularising, ensuring posterior propriety and numerical stability without being tuned to the proposed approximation.
We adopt the default prior families commonly used in Bayesian SEM software (as implemented in blavaan), assigning diffuse Gaussian priors to unconstrained coefficients (e.g., loadings and regression effects) and weakly informative priors to (co)variance components, with covariance structures parameterised via SDs and correlations.
All priors are specified independently across estimable parameters, and they are fixed a priori and held constant across the benchmark and subsequent simulation studies.
Table~\ref{tbl-priors} lists the diffuse priors used for the benchmark and simulation studies, alongside an informative prior specification used in the calibration analysis of Section~\ref{sec-sbc}.

\begin{table}[tbp]

\caption{\label{tbl-priors}Prior specification for the benchmark SEM. Priors are placed on the parameters as listed; the last column gives the corresponding element of \(\boldsymbol\vartheta\) on which the approximation operates, with the prior being carried across by change of variables. Beta priors for correlations are rescaled to \(\rho_{i,j}\in(-1,1)\).}

\centering{

\fontsize{10.0pt}{12.0pt}\selectfont
\begin{tabular*}{\linewidth}{@{\extracolsep{\fill}}lllll}
\toprule
 &  & \multicolumn{2}{c}{{Prior}} &  \\ 
\cmidrule(lr){3-4}
Parameter & Type & Diffuse & Informative & \(\boldsymbol\vartheta\) \\ 
\midrule\addlinespace[2.5pt]
$\nu_{i}$ & Observed intercept & $\N(0,32^2)$ & $\N(0,32^2)$ & $\nu_i$ \\ 
$\alpha_j$ & Latent intercept & $\N(0,10^2)$ & $\N(0,10^2)$ & $\alpha_j$ \\ 
$\Lambda_{i,j}$ & Loading & $\N(0,10^2)$ & $\N(1.25,0.25^2)$ & $\Lambda_{i,j}$ \\ 
$B_{j,j'}$ & Regression coefficient & $\N(0,10^2)$ & $\N(1.5,0.25^2)$ & $B_{j,j'}$ \\ 
$\Theta_{i,i}^{1/2}$ & Residual standard deviation & $\Gamma(1.0,0.5)$ & $\Gamma(10,10)$ & $\log \Theta_{i,i}$ \\ 
$\Psi_{j,j}^{1/2}$ & Latent standard deviation & $\Gamma(1.0,0.5)$ & $\Gamma(10,10)$ & $\log \Psi_{j,j}$ \\ 
$\rho_{i,j}$ & Correlations associated with covariances & $\text{Beta}(1,1)$ & $\text{Beta}(5,5)$ & $\operatorname{artanh} \rho_{i,j}$ \\ 
\bottomrule
\end{tabular*}

}

\end{table}%

All analyses were conducted in R version 4.6.1 \autocite{Rlang}.
The proposed INLA-based approximation was implemented in the INLAvaan package version 0.3.2 \autocite{jamil2026inlavaan}.
For MCMC, we depended on blavaan version 0.6-1 \autocite{merkle2021efficient}, which uses Stan's Hamiltonian Monte Carlo (HMC) sampler \autocite{standevelopmentteam2026stan}.
Computations were performed on a MacBook Pro (Apple M4 Pro, 14-core CPU, 24 GB unified RAM).

\subsection{MCMC Comparison}\label{sec-mcmc}

We assess the accuracy of the proposed deterministic approximation on the canonical benchmark model by comparison to a high-accuracy MCMC reference.
The assessment targets the full set of posterior marginals for the SEM parameters, with particular emphasis on uncertainty summaries and on the fidelity of the marginal density shape.
It then extends to further quantities recovered from the joint posterior sample, providing an additional and more demanding test of the copula construction in Section~\ref{sec-copula}.

For comparison, MCMC samples were obtained with three independent chains, 5,000 warmup iterations per chain, and 10,000 post-warmup draws per chain.
Convergence diagnostics indicated excellent mixing: \(\widehat R=1.00\) for all parameters and a minimum total effective sample size (ESS) of 9,048.
Each MCMC chain took roughly 23 seconds to run.

Our proposed INLA-based approximation completed inference in 1.43 seconds in total.
Breaking this down further, the joint Laplace approximation (mode-finding and Hessian forming) took 95 ms, the variational mean correction took 102 ms, the marginal construction took 901 ms for the 42 parameters in serial, the copula sampling and summarisation took 115 ms, and the fit statistics reported below took the remaining 219 ms.

For each parameter \(\vartheta_j\), we compute absolute errors between the approximation and MCMC for the posterior mean, SD, median, central 95\% interval endpoints \((q_{.025},q_{.975})\), and mode.
These errors are summarised by parameter class in Table~\ref{tbl-mcmc-compare}.
In addition, we report a standardised mean discrepancy,
\[
\Delta_{\mathrm{Mean},j}
:=
\frac{\left| \operatorname{E}_{\text{approx}}(\vartheta_j)-\operatorname{E}_{\text{MCMC}}(\vartheta_j)\right|}{\operatorname{SD}_{\text{MCMC}}(\vartheta_j)},
\]
which expresses mean error relative to posterior uncertainty.
Values well below 1 indicate approximation error that is small compared to posterior dispersion.

\begin{figure}[htbp]

\centering{

\includegraphics[width=1\linewidth,height=\textheight,keepaspectratio]{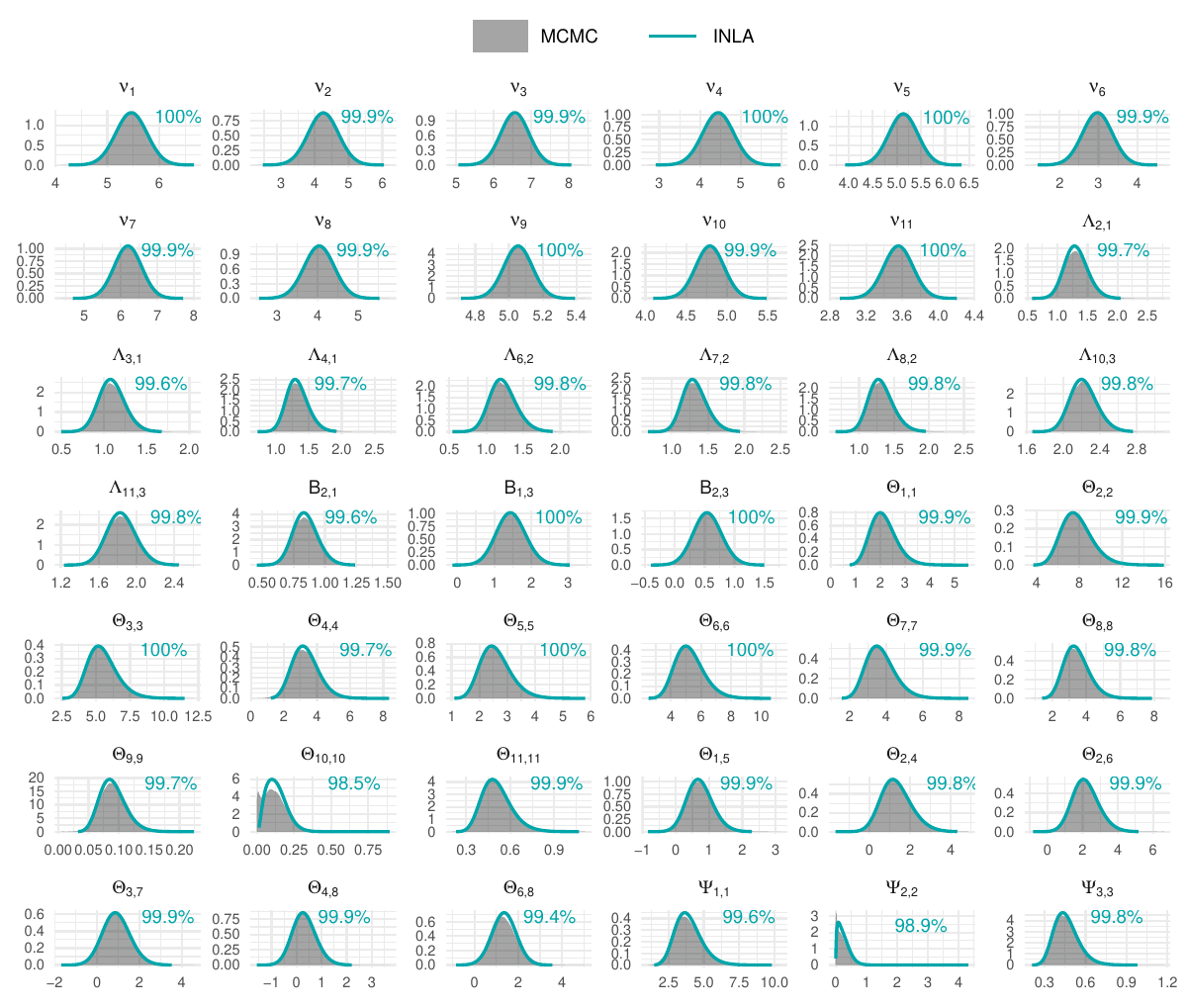}

}

\caption{\label{fig-mcmc-compare}Marginal posterior densities for the Political Democracy SEM. Fitted skew-normal approximations are overlaid on MCMC kernel density estimates. Percentages indicate JS similarity, with values near 100\% reflecting close distributional agreement.}

\end{figure}

To assess the agreement of full marginal shapes, we compute a symmetric divergence between the marginal densities implied by the approximation and MCMC.
Let \(\tilde\pi_j := \tilde\pi(\vartheta_j \mid \mathbf y)\) denote the fitted skew-normal approximation and let \(\pi_j\) denote a smooth density estimate from MCMC draws, which serves as the reference.
We report the Jensen--Shannon (JS) divergence
\[
D_{\mathrm{JS}}(\tilde\pi_j,\pi_j)
=
\tfrac12 D_{\mathrm{KL}}(\tilde\pi_j \ \Vert \ \bar\pi_j) + \tfrac12 D_{\mathrm{KL}}(\pi_j \ \Vert \ \bar\pi_j), \qquad \bar\pi_j=\tfrac12(\tilde\pi_j+\pi_j),
\]
and present results as a bounded ``percent similarity'' \(100\% \times \left\{ 1-  \frac{D_{\mathrm{JS}}(\tilde\pi_j,\pi_j)}{\log 2}\right\}\), where 100\% indicates identical distributions.
The JS divergence is invariant to monotone reparameterisation, so the scale on which it is computed is immaterial.
Figure~\ref{fig-mcmc-compare} overlays the fitted skew-normal approximation on the MCMC density for each marginal.

Overall, the proposed approximation shows strong agreement with the MCMC reference across all parameter classes (Table~\ref{tbl-mcmc-compare}).
Median standardised discrepancies \(\Delta_{\mathrm{Mean},j}\) are all below 0.07 across all parameter classes, and the single largest value observed across all 42 parameters is 0.23 posterior SDs---well within the range that would leave substantive inference unaffected.

Intercepts, loadings, and regression coefficients are recovered with negligible error.
Mean absolute errors for posterior means and central quantiles are of the order of 0.01 or smaller, as expected given that these parameters typically yield approximately symmetric, well-behaved posteriors.
Somewhat larger discrepancies arise for variance and covariance parameters, particularly in the tail quantiles (\(q_{.025}\), \(q_{.975}\)), reflecting the limited flexibility of the skew-normal family in capturing the shape of distributions that are bounded below at zero and often exhibit pronounced right skew.
The residual covariances are a special case among these, in that they are the only quantities in the table not obtained by profiling.
A covariance \(\Theta_{i,j} = \rho_{i,j}(\Theta_{i,i}\Theta_{j,j})^{1/2}\) is a function of three estimated parameters at once and so admits no axis of its own to scan, and it is summarised instead from the copula sample of Section~\ref{sec-copula}.
It is seen that the errors remain small in absolute terms even for these components, especially for posterior means and medians, which are the summaries most commonly used in applied SEM inference.

The density overlays in Figure~\ref{fig-mcmc-compare} corroborate these findings.
JS similarity exceeds 99.6\% for the majority of parameters, and the minimum across all 42 marginals is 98.5\% (\(\Theta_{10,10}\)), confirming close distributional agreement throughout.
Inspection of the cases with the lowest similarity reveals their sources.
Variance marginals (e.g., \(\Theta_{10,10}\) and \(\Psi_{2,2}\)) exhibit an upward inflection of the MCMC density near zero, consistent with the sampler encountering the positivity boundary---a hallmark of near-Heywood conditions.
A unimodal skew-normal density cannot reproduce this boundary pile-up, yet it compensates by concentrating mass in the region of highest posterior probability, preserving the fidelity of central summaries.
The latent variance \(\Psi_{2,2}\) displays a notably elongated right tail in the MCMC reference, suggestive of weak identification or insufficient prior regularisation.
While the skew-normal approximation captures the bulk of this distribution, it under-represents the extreme tail, as expected.

\setlength{\tabcolsep}{4.5pt}

\begin{table}[tbp]

\caption{\label{tbl-mcmc-compare}Mean absolute error (MAE) of posterior summary quantities, aggregated by parameter class. \emph{Max (Mean)} reports the maximum absolute error of the posterior mean within each class, \(\Delta_{\mathrm{Mean},j}\) denotes the standardised discrepancy, and \(N\) is the parameter count per class.}

\centering{

\fontsize{10.0pt}{12.0pt}\selectfont
\begin{tabular*}{\linewidth}{@{\extracolsep{\fill}}l|rrrrrrr>{\raggedleft\arraybackslash}p{\dimexpr 37.50pt -2\tabcolsep-1.5\arrayrulewidth}rr}
\toprule
 &  & \multicolumn{6}{c}{{MAE (absolute)}} &  & \multicolumn{2}{c}{{\(\Delta_{\mathrm{Mean},j}\)}} \\ 
\cmidrule(lr){3-8} \cmidrule(lr){10-11}
Parameter & \(N\) & Mean & SD & \(q_{.025}\) & \(q_{.50}\) & \(q_{.975}\) & Mode & Max (Mean) & Median & Max \\ 
\midrule\addlinespace[2.5pt]
\multicolumn{1}{l}{{Observed intercepts}} & 11 & 0.003 & 0.009 & 0.024 & 0.003 & 0.014 & 0.018 & 0.006 & 0.010 & 0.017 \\ 
\multicolumn{1}{l}{{Loadings}} & 8 & 0.010 & 0.016 & 0.013 & 0.007 & 0.053 & 0.011 & 0.013 & 0.052 & 0.075 \\ 
\multicolumn{1}{l}{{Regressions}} & 3 & 0.005 & 0.008 & 0.014 & 0.006 & 0.016 & 0.034 & 0.009 & 0.021 & 0.039 \\ 
\multicolumn{1}{l}{{Residual variances}} & 11 & 0.012 & 0.024 & 0.040 & 0.012 & 0.062 & 0.038 & 0.024 & 0.025 & 0.225 \\ 
\multicolumn{1}{l}{{Residual covariances}} & 6 & 0.017 & 0.029 & 0.028 & 0.019 & 0.112 & 0.046 & 0.052 & 0.012 & 0.068 \\ 
\multicolumn{1}{l}{{Latent variances}} & 3 & 0.027 & 0.031 & 0.075 & 0.032 & 0.047 & 0.046 & 0.045 & 0.045 & 0.164 \\ 
\multicolumn{1}{l}{{\bfseries Overall}} & {\bfseries 42} & {\bfseries 0.011} & {\bfseries 0.019} & {\bfseries 0.030} & {\bfseries 0.011} & {\bfseries 0.050} & {\bfseries 0.029} & {\bfseries 0.052} & {\bfseries 0.019} & {\bfseries 0.225} \\ 
\bottomrule
\end{tabular*}
\begin{minipage}{\linewidth}
\vspace{.05em}
MCMC: 3 chains, 10,000 post-warmup draws each.\\
\end{minipage}

}

\end{table}%

\setlength{\tabcolsep}{6pt}

\protect\phantomsection\label{cell-fig-mcmc-compare}

\begin{figure}[htbp]

\centering{

\includegraphics[width=1\linewidth,height=\textheight,keepaspectratio]{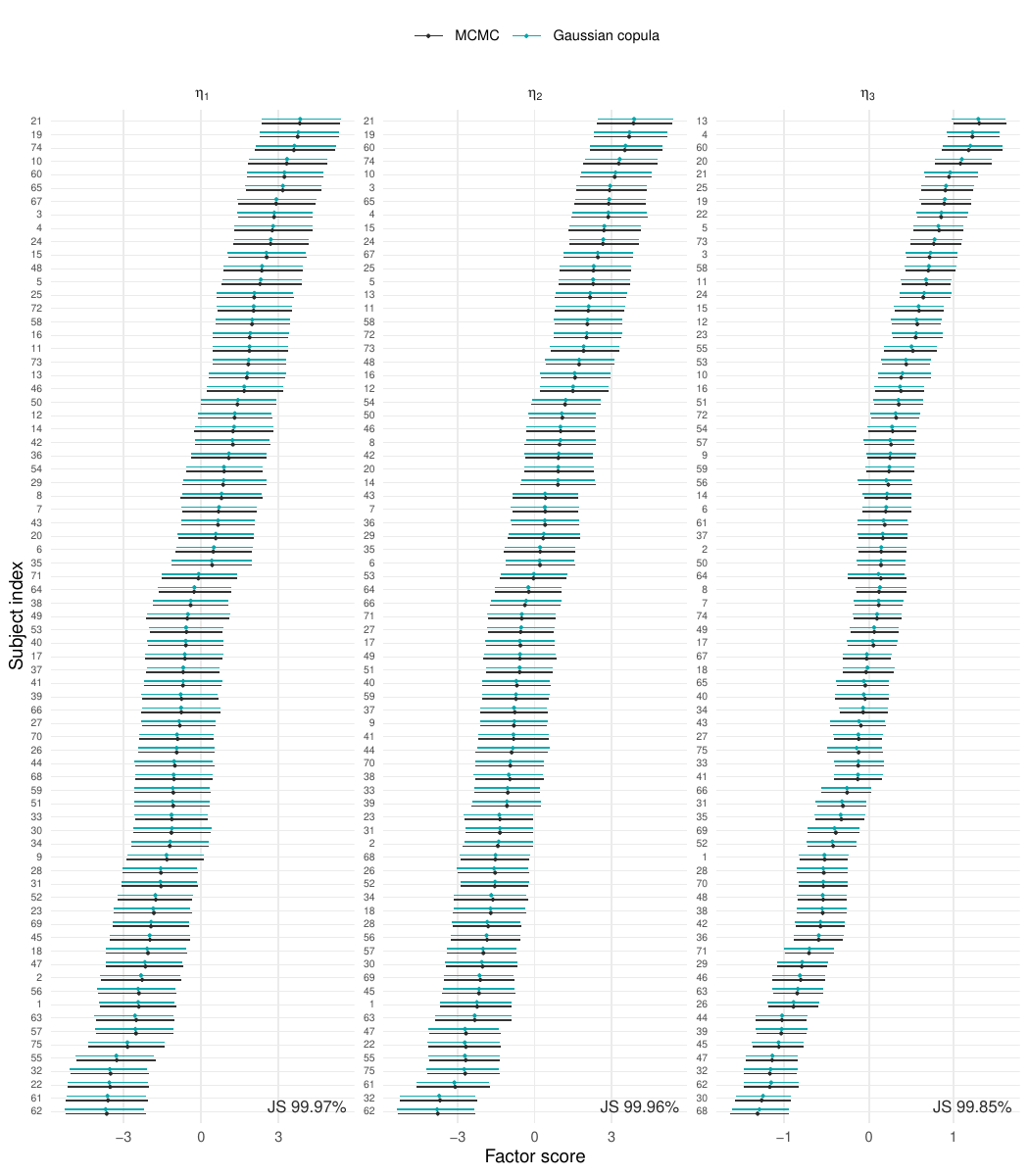}

}

\caption{\label{fig-factorscores}Posterior factor scores for all 75 individuals in the Political Democracy data, by latent variable. Points are posterior means and bars the central 95\% credible intervals, comparing the MCMC reference against the Gaussian copula sample of Section~\ref{sec-copula}. Within each panel, subjects are ordered by their MCMC posterior mean. Annotations give the median JS similarity across the 75 cases, with 100\% indicating identical distributions.}

\end{figure}

Turning our attention to factor scores provides a more exacting test of the reconstruction.
Recovering the latent variables entails drawing \(\boldsymbol\eta_s \sim \N_q\big(\mathbf m_s(\boldsymbol\vartheta),\mathbf V(\boldsymbol\vartheta)\big)\) from (\ref{eq-cond-post-eta}) at each copula draw of \(\boldsymbol\vartheta\).
The resulting factor-score posterior is a mixture over the reconstructed joint posterior and is sensitive to discrepancies in both its marginals and dependence structure.
Both estimation approaches use the same conditional distribution for \(\boldsymbol\eta_s\), so Figure~\ref{fig-factorscores} isolates the effect of replacing MCMC draws of \(\boldsymbol\vartheta\) with copula draws.
Neither the locations nor the widths of the two sets of credible intervals differ systematically.
Across the 75 cases, the median JS similarities are 99.97\%, 99.96\%, and 99.85\% for \(\eta_1\), \(\eta_2\), and \(\eta_3\), with corresponding standardised mean discrepancies of 0.020, 0.023, and 0.043 MCMC posterior SDs.

For completeness, we also checked the model-fit indices produced from the same joint copula draws, which showed similarly close agreement.
We computed a posterior predictive \(p\)-value \autocite{gelman1996posterior} of 0.507 against 0.505 for MCMC.
The deviance information criterion \autocite[DIC;][]{spiegelhalter2002bayesian} is 3,179 under both, the approximation trading a lower mean deviance (3,137.9 against 3,138.7) for a larger effective parameter count (40.9 against 39.8).

\protect\phantomsection\label{cell-fig-factorscores}

\subsection{Error Ablation of the Marginal Construction}\label{sec-ablation}

While the preceding comparison validates the estimation framework in aggregate, it would be useful, by way of an ablation study, to disentangle the individual contributions of the stacked approximations that produce it.
We therefore repeat the comparison against three constructions of the marginals, in the order in which the method constructs them: (i) those implied by the mode-centred joint Laplace approximation, which are Gaussian, \(\N(\vartheta_j^*, \Omega_{jj})\); (ii) the same Gaussians recentred by the VB mean shift of Section~\ref{sec-vb} at \(\vartheta_j^* + \hat\delta_j\); and (iii) the volume-corrected skew-normal marginals of Section~\ref{sec-marginals}.
The first two differ only in where the same family of marginals is placed, while the third replaces that family outright.
All three fits share the same posterior mode and Hessian, so the comparison isolates the post-optimisation layers.
So that every stage yields a deterministic marginal for each parameter, the six residual covariances are evaluated as the correlations \(\rho_{i,j}\) through which they are parameterised (Table~\ref{tbl-priors}), with the MCMC reference obtained by transforming each posterior draw as \(\rho_{i,j} = \Theta_{i,j} \big/ \big(\Theta_{i,i}\Theta_{j,j}\big)^{1/2}\).
None of the three fits therefore rely on posterior sampling, leaving the copula out of the comparison.

Averaged over all 42 parameters, both metrics improve at every stage (Table~\ref{tbl-ablation-ladder}).
The VB shift reduces the mean JS discrepancy (the complement of JS similarity) from 1.34\% to 0.50\% and the standardised mean discrepancy from 0.19 to 0.04 posterior SDs, confirming that at this sample size the gap between posterior mode and mean is the dominant error term.
The skew-normal marginals cut the JS discrepancy further, to 0.19\%, since the fitted skew-normal captures the full asymmetric shape of each marginal, but reduce the mean discrepancy only marginally, to 0.035.
Relative to the bare Laplace approximation, the skew-normal construction reduces the JS discrepancy more than sevenfold and the standardised mean discrepancy roughly fivefold, at a cost of just one additional second of computation.

\begin{table}[tbp]

\caption{\label{tbl-ablation-ladder}Error ablation of the marginal construction on the benchmark model, by parameter class. The JS discrepancy is the complement of the JS similarity (smaller is better). \(\Delta_{\mathrm{Mean},j}\) denotes the standardised discrepancy, and \(N\) denotes the number of parameters in each class.}

\centering{

\fontsize{10.0pt}{12.0pt}\selectfont
\begin{tabular*}{\linewidth}{@{\extracolsep{\fill}}l|rrrrrrr}
\toprule
 &  & \multicolumn{3}{c}{{JS discrepancy (\%)}} & \multicolumn{3}{c}{{\(\Delta_{\mathrm{Mean},j}\)}} \\ 
\cmidrule(lr){3-5} \cmidrule(lr){6-8}
Parameter & \(N\) & (i) & (ii) & (iii) & (i) & (ii) & (iii) \\ 
\midrule\addlinespace[2.5pt]
\multicolumn{1}{l}{{Observed intercepts}} & 11 & 0.05 & 0.05 & 0.05 & 0.011 & 0.011 & 0.011 \\ 
\multicolumn{1}{l}{{Loadings}} & 8 & 2.40 & 0.89 & 0.24 & 0.311 & 0.086 & 0.053 \\ 
\multicolumn{1}{l}{{Regressions}} & 3 & 0.54 & 0.28 & 0.17 & 0.127 & 0.012 & 0.023 \\ 
\multicolumn{1}{l}{{Residual variances}} & 11 & 2.42 & 0.64 & 0.24 & 0.317 & 0.058 & 0.044 \\ 
\multicolumn{1}{l}{{Residual correlations}} & 6 & 0.28 & 0.11 & 0.10 & 0.096 & 0.027 & 0.022 \\ 
\multicolumn{1}{l}{{Latent variances}} & 3 & 2.25 & 1.64 & 0.56 & 0.278 & 0.037 & 0.082 \\ 
\multicolumn{1}{l}{{\bfseries Overall}} & {\bfseries 42} & {\bfseries 1.34} & {\bfseries 0.50} & {\bfseries 0.19} & {\bfseries 0.188} & {\bfseries 0.042} & {\bfseries 0.035} \\ 
\bottomrule
\end{tabular*}
\begin{minipage}{\linewidth}
\vspace{.05em}
Constructions: (i) Gaussian marginals, mode-centred; (ii) Gaussian marginals, VB-centred; (iii) skew-normal marginals, volume-corrected (the default). Total wall-clock time 0.21, 0.38 and 1.12 s respectively.\\
\end{minipage}

}

\end{table}%

The largest gains in Table~\ref{tbl-ablation-ladder} occur for loadings, residual variances, and latent variances, whose marginals depart furthest from Gaussianity, with initial JS discrepancies of approximately 2.3--2.4\%.
While the skew-normal construction improves overall distributional agreement, it need not improve mean accuracy in every parameter class.
For regressions and latent variances, (ii) estimates posterior means slightly more accurately than (iii) (0.01 vs.~0.02 and 0.04 vs.~0.08 SDs, respectively).
Notably, the difference for variances is driven by \(\Psi_{2,2}\), which was also among the more challenging marginals in the MCMC benchmark (Section~\ref{sec-mcmc}).
Here, (ii) happens to match its natural-scale mean more closely, although (iii) better captures its marginal shape.

The JS discrepancy for the 11 observed intercepts remains at 0.05\% throughout.
Under the benchmark's saturated mean structure, \(\boldsymbol\mu(\boldsymbol\vartheta) = \boldsymbol\nu\), the intercepts enter the log-posterior quadratically and, with diffuse Gaussian priors, are essentially information-orthogonal to covariance parameters at the mode.
Their marginals are symmetric scale mixtures of Gaussians about a common centre, so the variational shift is negligible and later layers offer no improvement.
This symmetry argument relies on the benchmark's saturated mean structure and need not extend to more general mean specifications such as that in {\eqref{eq-reduced-form-a}}.

Construction (iii) rests on the efficient volume correction of Corollary~\ref{cor-eff-vol}, which is evaluated afresh at every scan direction and whose cost therefore grows with the dimension of the model.
Two questions naturally follow: what the economy costs, and what the correction itself is worth.
Replacing the gradient-only scheme with a reference implementation that forms the full numerical Hessian leaves both metrics unchanged at the reported precision, 0.19\% and 0.035 in both cases, while lengthening the fit by 0.8 seconds.
The shortcut thus reaches the same marginals sooner at no measurable cost in accuracy.
Dispensing with the correction altogether is another matter.
Effectively, what remains is a first-order construction that ignores the tilt carrying the marginal's location, and the mean JS discrepancy rises in aggregate to 0.91\%, exceeding even the Gaussian marginals of (ii).
The per-class results can be inspected in the appendix (Table~\ref{tbl-gamma1-appendix}).

\subsection{Simulation Study}\label{sec-simulation}

We complement the single-dataset comparison with a repeated-sampling study that evaluates the operating conditions of the proposed approximation.
Our method relies on local Laplace expansions for both the joint posterior (Section~\ref{sec-vb}) and the marginal posteriors (Section~\ref{sec-marginals}).
Downstream steps (namely, the variational mean correction and the copula reconstruction of Section~\ref{sec-copula}) use these expansions to produce the final posterior summaries, so inaccuracies in either expansion may propagate to all subsequent inferences.
Guided by the regularity conditions in Section~\ref{sec-preliminaries}, we systematically experiment with posterior concentration, local identifiability, and boundary proximity, together with model dimension, in order to identify the data and model conditions under which our approximation remains reliable.

\begin{figure}[htbp]

\centering{

\includegraphics[width=0.97\linewidth,height=\textheight,keepaspectratio]{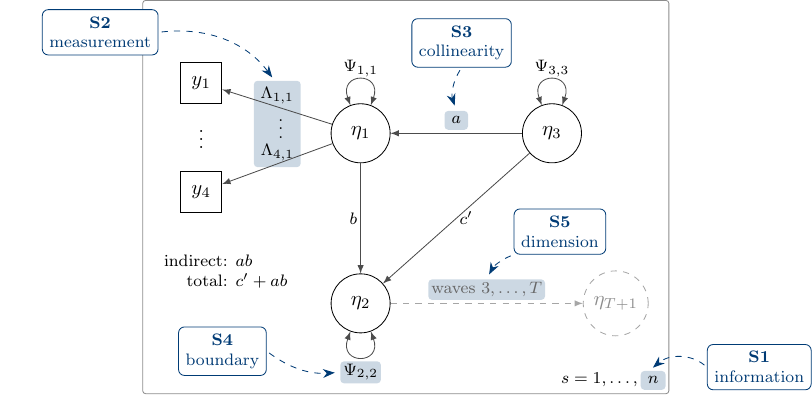}

}

\caption{\label{fig-sentinel}Simulation design mapped onto the Political Democracy benchmark (Figure~\ref{fig-poldem}), with highlighted components indicating the perturbations specified in Table~\ref{tbl-stress-map}. Only the indicator block of \(\eta_1\) is shown.}

\end{figure}%

Rather than survey a model zoo of SEMs, we make five targeted modifications to the Political Democracy benchmark, taking as the true parameter vector \(\boldsymbol\vartheta_0\) a bias-reduced regularised maximum likelihood estimate \autocite{jamil2026biasreduced}.
The first four scenarios probe these requirements, while the fifth varies the dimension of the model; Figure~\ref{fig-sentinel} depicts each modification, and Table~\ref{tbl-stress-map} summarises the design.
Scenario S1 varies the sample size, and with it the degree of posterior concentration.
Scenarios S2--S4 modify the measurement model of \(\eta_1\) and the structural relations among the factors, which form a latent partial mediation system with paths \(a\) (\(\eta_3 \to \eta_1\)), \(b\) (\(\eta_1 \to \eta_2\)), and \(c'\) (\(\eta_3 \to \eta_2\)).
Scenario S2 lowers the standardised loadings on \(\eta_1\) from 0.80 to 0.30, weakening the information the indicators carry about the factor.
Scenario S3 raises the predictor overlap \(R^2_{13} = \operatorname{Cor}(\eta_1, \eta_3)^2\) from 0.20 to 0.98, so that \(\eta_1\) and \(\eta_3\), the two predictors of \(\eta_2\), grow increasingly collinear and \(b\) and \(c'\) increasingly hard to separate.
Scenario S4 reduces the disturbance variance of \(\eta_2\) to 1\% of the factor's total variance, driving \(\Psi_{2,2}\) towards the boundary of the parameter space.
Scenario S5 extends the panel from two to four and eight waves (\(m \in \{42, 80, 156\}\) free parameters).
Each additional wave \(\tau = 3, \dots, T\) contributes a four-indicator factor \(\eta_{\tau+1}\), regressed on \(\eta_3\) and the previous wave, proportionately adding indicators and correlated residuals, with generating values for the added waves replicating the second-wave block.
By varying measurement, structure, and dimension within a common data-generating framework, these scenarios reproduce in controlled fashion the difficulties that arise in confirmatory factor, latent mediation, and longitudinal models.

\begin{table}[tbp]

\caption{\label{tbl-stress-map}Targeted scenarios for the simulation study. Underlined values are those of the baseline model using \(\boldsymbol\vartheta_0\). S2--S4 are evaluated at \(n = 75\); for S2, the stated loading value is for all indicators of \(\eta_1\) simultaneously (by rescaling their residual variances). For S3, \(\operatorname{Var}(\eta_1)\) is held at its baseline value throughout.}

\centering{

\fontsize{10.0pt}{12.0pt}\selectfont
\begin{tabular*}{\linewidth}{@{\extracolsep{\fill}}lll}
\toprule
Scenario & Conditions & Feature examined \\ 
\midrule\addlinespace[2.5pt]
S1: Information & $n \in \{40, \underline{75}, 150, 300, 600, 1200\}$ & Posterior concentration \\ 
S2: Measurement & $\boldsymbol\lambda^{\mathrm{std}}_{\eta_1} \in \{\underline{\approx\! 0.80}, 0.50, 0.30\}$ & Identification of $\eta_1$ \\ 
S3: Collinearity & $R^2_{13} = \operatorname{Cor}(\eta_1, \eta_3)^2 \in \{\underline{0.20}, 0.80, 0.95, 0.98\}$ & Identification of $b$ and $c'$ \\ 
S4: Boundary & $\Psi_{2,2}/\operatorname{Var}(\eta_2) \in \{0.50, 0.15, \underline{0.045}, 0.01\}$ & Pull of $\Psi_{2,2}$ towards zero \\ 
S5: Dimension & $T \in \{\underline{2}, 4, 8\}$ $\big(m \in \{\underline{42}, 80, 156\}\big)$, $n \in \{\underline{75}, 300\}$ & Scaling with parameter dimension \\ 
\bottomrule
\end{tabular*}

}

\end{table}%

Varying one quantity at a time about the shared baseline at \(\boldsymbol\vartheta_0\) (\(n = 75\)), and counting shared conditions only once, gives 18 conditions: 14 for S1--S4 and four unique to S5.
Each condition comprises 100 replications under the diffuse priors of Table~\ref{tbl-priors}, with replications of the same index sharing a seed throughout, so as to isolate the effect of the experimental settings from sampling variability.
Every replication is fitted once by the proposed approximation, and once more in blavaan using three MCMC chains with 5,000 warmup and 2,000 retained draws per chain (retaining a fit only if \(\max_j \widehat R_j < 1.05\) and \(\min_j \mathrm{ESS}_j > 100\) across all \(m\) parameters).
The paired MCMC fit serves as the reference against which the standardised mean discrepancy \(\Delta_{\mathrm{Mean},j}\) and JS similarity of Section~\ref{sec-mcmc} are computed.

Summaries for S1--S4 are computed on the same set of 44 quantities, the 42 free parameters of the baseline model together with the derived effects \(ab\) and \(c' + ab\).
For S5 the set is every free parameter of the fitted panel together with the same two effects, so the larger models are searched in full.
Against the paired reference we report the per-replication worst case, i.e., the largest standardised mean discrepancy over the set and the smallest JS similarity, summarised across replications.
Taking the maximum avoids dilution across quantities, and we record which quantity contributes the largest discrepancy under each condition.

We also report 95\% credible-interval coverage of the generating value (CR95) over the set, together with two fit diagnostics, the VB shift \(\hat{\boldsymbol\delta}\) of Section~\ref{sec-vb}, and the normalised maximum absolute deviation (NMAD), which quantifies the fidelity of the fitted skew-normal curve to the scan it approximates.
For parameter \(j\), with volume-corrected ordinates \(h_{kj}\) as in (\ref{eq-corrected-ordinates}) and fitted parameters \((\hat\xi_j,\hat\omega_j,\hat\alpha_j,\hat c_j)\),
\[
\mathrm{NMAD}_j
=
\frac{\max_k \big| \exp(h_{kj}) - \exp\{\log f_{\mathrm{SN}}(\vartheta_j(t_k);\hat\xi_j,\hat\omega_j,\hat\alpha_j)+\hat c_j\}\big|}{\max_k \exp(h_{kj})},
\]
the largest exponentiated fitting residual as a fraction of the peak profile ordinate, so that \(\mathrm{NMAD}_j = 0\) denotes exact agreement at every scanned point.

Table~\ref{tbl-operating} collects the results.
Read row by row, the table separates into two regimes.
Along the information, boundary, and dimension scenarios the maximum discrepancy lies between 0.07 and 0.32 SDs, the minimum JS similarity stays above 92\%, and coverage stays within about three points of nominal, which is the accuracy of the baseline at \(n = 75\).
Under weak measurement the discrepancy rises to 0.69 and 1.63 SDs, the minimum JS similarity falls to 44\%, and only 36 of the 100 references remain usable, while severe collinearity produces a milder version of the same pattern, with discrepancies of 0.30 to 0.35 and similarities in the 82--87\% range.
Overall, the loss of accuracy is driven by identification, as opposed to information, boundary, or dimension.
Weak measurement is the severe case and collinearity the milder one, and in both the harm concentrates in a few quantities while the rest of the model is recovered at baseline accuracy, as the following paragraphs elaborate.

\setlength{\tabcolsep}{4pt}

\begin{table}[tbp]

\caption{\label{tbl-operating}Results of the simulation study, summarising the 18 conditions across all \(m\) free parameters of the fitted model together with the derived effects \(ab\) and \(c' + ab\). CR95 is coverage of the generating truth over the set. \emph{NMAD} and \emph{VB shift} are per-fit maxima, summarised across replications as median (95th percentile). The \emph{vs.~MCMC} columns are errors against the paired MCMC reference, including the per-fit maximum standardised mean discrepancy, as median (95th percentile); the per-fit minimum JS similarity, as median; \emph{Ref.}, the number of usable references out of the 100 replications paired with MCMC; and \emph{Worst}, the quantity with the largest discrepancy in a replication, reported as the most frequent such quantity together with its share of the usable replications.}

\centering{

\fontsize{9.0pt}{11.0pt}\selectfont
\begin{tabular*}{\linewidth}{@{\extracolsep{\fill}}>{\raggedright\arraybackslash}p{\dimexpr 86.25pt -2\tabcolsep-1.5\arrayrulewidth}>{\raggedleft\arraybackslash}p{\dimexpr 36.00pt -2\tabcolsep-1.5\arrayrulewidth}rrrrr>{\raggedright\arraybackslash}p{\dimexpr 36.00pt -2\tabcolsep-1.5\arrayrulewidth}>{\raggedleft\arraybackslash}p{\dimexpr 36.00pt -2\tabcolsep-1.5\arrayrulewidth}}
\toprule
 &  & \multicolumn{2}{c}{{Diagnostics}} & \multicolumn{5}{c}{{vs. MCMC}} \\ 
\cmidrule(lr){3-4} \cmidrule(lr){5-9}
Condition & CR95 (\%) & NMAD & VB shift & \(\max_j\!\) \(\Delta_{\mathrm{Mean},j}\) & \(\min_j\!\) JS (\%) & Ref. & Worst & Share (\%) \\ 
\midrule\addlinespace[2.5pt]
\multicolumn{9}{l}{{\itshape S1: Information}} \\[2.5pt] 
\midrule\addlinespace[2.5pt]
$n = 1200$ & 95.2 & 0.023 (0.029) & 0.24 (0.46) & 0.07 (0.25) & 99.3 & 100 & $\Psi_{2,2}$ & 42 \\ 
$n = 600$ & 94.5 & 0.029 (0.032) & 0.31 (0.55) & 0.12 (0.26) & 97.8 & 100 & $\Psi_{2,2}$ & 66 \\ 
$n = 300$ & 94.1 & 0.031 (0.041) & 0.40 (0.71) & 0.20 (0.28) & 96.7 & 99 & $\Psi_{2,2}$ & 72 \\ 
$n = 150$ & 94.1 & 0.037 (0.047) & 0.43 (0.67) & 0.21 (0.33) & 96.2 & 90 & $\Psi_{2,2}$ & 53 \\ 
{\bfseries $\boldsymbol{n = 75}$} & {\bfseries 92.4} & {\bfseries 0.041 (0.054)} & {\bfseries 0.51 (0.69)} & {\bfseries 0.23 (0.32)} & {\bfseries 95.5} & {\bfseries 70} & {\bfseries $\boldsymbol{\Theta_{10,10}}$} & {\bfseries 71} \\ 
$n = 40$ & 92.3 & 0.049 (0.065) & 0.59 (0.81) & 0.32 (0.74) & 92.9 & 47 & $\Lambda_{4,1}$ & 21 \\ 
\midrule\addlinespace[2.5pt]
\multicolumn{9}{l}{{\itshape S2: Measurement}} \\[2.5pt] 
\midrule\addlinespace[2.5pt]
loadings 0.5 & 92.0 & 0.076 (0.100) & 0.62 (0.75) & 0.69 (1.37) & 76.6 & 48 & $\Psi_{1,1}$ & 75 \\ 
loadings 0.3 & 87.9 & 0.102 (0.129) & 0.61 (0.89) & 1.63 (3.73) & 43.5 & 36 & $\Psi_{1,1}$ & 81 \\ 
\midrule\addlinespace[2.5pt]
\multicolumn{9}{l}{{\itshape S3: Collinearity}} \\[2.5pt] 
\midrule\addlinespace[2.5pt]
$R^2 = 0.80$ & 93.3 & 0.043 (0.054) & 0.46 (0.59) & 0.21 (0.27) & 95.8 & 85 & $\Psi_{2,2}$ & 56 \\ 
$R^2 = 0.95$ & 93.2 & 0.046 (0.071) & 0.59 (1.00) & 0.30 (0.66) & 87.0 & 78 & $\Psi_{1,1}$ & 49 \\ 
$R^2 = 0.98$ & 93.5 & 0.048 (0.081) & 0.70 (1.08) & 0.35 (0.71) & 82.8 & 80 & $\Psi_{1,1}$ & 50 \\ 
\midrule\addlinespace[2.5pt]
\multicolumn{9}{l}{{\itshape S4: Boundary}} \\[2.5pt] 
\midrule\addlinespace[2.5pt]
fraction 0.50 & 92.5 & 0.039 (0.054) & 0.48 (0.75) & 0.24 (0.64) & 95.0 & 65 & $\Theta_{10,10}$ & 49 \\ 
fraction 0.15 & 93.6 & 0.036 (0.054) & 0.47 (0.72) & 0.24 (0.33) & 95.4 & 72 & $\Theta_{10,10}$ & 72 \\ 
fraction 0.01 & 92.5 & 0.043 (0.053) & 0.55 (0.73) & 0.23 (0.32) & 94.9 & 76 & $\Theta_{10,10}$ & 67 \\ 
\midrule\addlinespace[2.5pt]
\multicolumn{9}{l}{{\itshape S5: Dimension}} \\[2.5pt] 
\midrule\addlinespace[2.5pt]
$m = 80$, $n = 75$ & 93.9 & 0.048 (0.055) & 0.59 (0.79) & 0.20 (0.33) & 95.6 & 55 & $\Theta_{10,10}$ & 49 \\ 
$m = 80$, $n = 300$ & 94.0 & 0.042 (0.050) & 0.43 (0.71) & 0.19 (0.25) & 96.7 & 94 & $\Psi_{2,2}$ & 39 \\ 
$m = 156$, $n = 75$ & 94.0 & 0.049 (0.058) & 0.64 (0.86) & 0.19 (0.28) & 95.3 & 47 & $\mathrm{Var}(\eta_{\tau=8})$ & 23 \\ 
$m = 156$, $n = 300$ & 94.5 & 0.045 (0.053) & 0.52 (0.83) & 0.20 (0.26) & 96.6 & 91 & $\Psi_{2,2}$ & 38 \\ 
\bottomrule
\end{tabular*}
\begin{minipage}{\linewidth}
\vspace{.05em}
The bold row is the benchmark at \(n = 75\). All INLA-based approximations converged with a positive-definite Hessian. MCMC: 3 chains, 5,000 warmup and 2,000 retained draws each; a reference is usable if all \(\widehat R < 1.05\) and the minimum ESS exceeds 100.\\
\end{minipage}

}

\end{table}%

\setlength{\tabcolsep}{6pt}

With increasing information, the estimation error decreases monotonically in \(n\), as the Bernstein--von Mises argument of Section~\ref{sec-preliminaries} predicts.
The median absolute error over all parameters falls from 0.28 at \(n = 40\) to 0.05 at \(n = 1200\), a factor of 5.6 over a thirtyfold increase in \(n\), close to the factor of 5.5 that the \(n^{-1/2}\) rate implies, and the maximum discrepancy against the reference falls from 0.32 to 0.07 SDs over the same range.
Small samples affect the reference before they affect the approximation.
At \(n = 40\), where \(n\) is smaller than \(m\), only about half of the MCMC references are usable, though we expect this would improve with stronger priors---an effect we investigate next in Section~\ref{sec-sbc}.
The approximation itself remains within about 0.3 SDs of the references that are usable.

Scenario S4 (Boundary) matches the baseline MCMC summaries well even when \(\Psi_{2,2}\) is reduced to 1\% of its factor variance, and the marginal of \(\Psi_{2,2}\) itself is recovered to within 0.08 SDs with a JS similarity of 96.7\%.
However, the approximation misses the lower tail, with lower tail mass \(\Pr(\Psi_{2,2} < 0.05 \times (\boldsymbol\Psi_0)_{2,2} \mid \mathbf y)\) of only 0.02\% (vs.~6.6\% for MCMC), reflecting the boundary pile-up that a unimodal skew-normal cannot capture (Section~\ref{sec-mcmc}).
The high-dimension conditions (\(m = 80\) and \(m = 156\)) similarly maintain this accuracy throughout.
Increasing dimension imposes only a modest computational cost, with median fit times (under parallel profiling) of 1.5, 5.6, and 9 seconds at \(m = 42\), 80, and 156, respectively, compared to per-chain medians of 15 seconds, 1 minute, and 8 minutes for the paired blavaan fits.

The approximation degrades where identification weakens, and the degradation appears in the quantity that the model structure predicts.
In the case of weak measurement, lowering the loadings of \(\eta_1\) to 0.3 pushes the worst-case discrepancy to 1.6 SDs.
The error is absorbed not by a loading, but predominantly by \(\Psi_{1,1}\) (in 81\% of replications) and the path leading into that factor, as the structural equation for \(\eta_1\) ties these moving quantities together.
This points to a fundamental data deficit, with two-thirds of the paired HMC fits failing here.
Even when both methods return a fit, they can disagree because the posterior is nearly flat along the trade-off between \(\Psi_{1,1}\) and the loadings of \(\eta_1\).
In any case, no method, sampling-based or otherwise, can recover curvature the data do not supply.

Conversely, in the case of collinearity between the predictors of \(\eta_2\), the loss is subtler and target-specific.
The mediation components lose shape fidelity together, with the JS similarity of \(b\), \(c'\), and \(ab\) falling to 84--85\% at \(R^2 = 0.98\), while their sum \(c' + ab\) retains 98\%.
Because \(ab\) is reconstructed through the copula of Section~\ref{sec-copula}, this scenario also serves as an audit of the copula-derived quantities.

To summarise this simulation study, the approximation can be expected to match MCMC whenever (a) the data carry enough information relative to \(m\) and (b) the parameters are separately identified in both the measurement and structural models.
It can be expected to fail, and to fail gradually, when identification weakens, first in the variance of a weakly measured factor and then in the components of a collinear decomposition, with the sampling-based reference failing alongside it.
A variance component pulled towards zero, a boundary the working parameterisation cannot cross, leaves the central summaries intact and affects only the tail probabilities of that variance.
Table~\ref{tbl-operating} indicates that the fit-time diagnostics track this degradation.
Specifically, the NMAD residual rises from 0.02--0.05 in benign conditions to 0.10 under weak measurement, while the VB shift peaks at 0.70 under severe collinearity, suggesting that expected error can be calibrated against these metrics.
We report such a calibration in the Supplementary Material.

\subsection{Simulation-Based Calibration Checking}\label{sec-sbc}

As a further internal consistency check, we perform simulation-based calibration checking \autocites[SBC;][]{talts2020validating,modrak2025simulationbased}.
SBC exploits the self-consistency identity
\[
\iint \pi(\boldsymbol{\vartheta} \mid \mathbf y)\, \pi(\mathbf y \mid \boldsymbol{\vartheta}^{\mathrm{samp}})\, \pi(\boldsymbol{\vartheta}^{\mathrm{samp}})\, d \mathbf y\, d \boldsymbol{\vartheta}^{\mathrm{samp}} = \pi(\boldsymbol{\vartheta}).
\]
If the inference algorithm is exact, then averaging the posterior over datasets drawn from the prior predictive distribution recovers the prior.
To operationalise this, one repeatedly (a) samples a parameter vector \(\boldsymbol{\vartheta}^{\mathrm{samp}}\) from the prior, (b) generates data \(\mathbf y \sim \pi(\mathbf y \mid \boldsymbol{\vartheta}^{\mathrm{samp}})\), and (c) fits the model to \(\mathbf y\).
For each replication and parameter \(\vartheta_j\), we compute the probability integral transform (PIT) value \(F_{\mathrm{SN}}^{(j)}(\vartheta_j^{\mathrm{samp}} \mid \mathbf y)\) using the CDF of the fitted skew-normal marginal.
The plots below label parameters in the natural SEM parameterisation of Table~\ref{tbl-priors}, which leaves the PITs unchanged because they are invariant under monotone transformations.
PITs for covariances use the CDFs of the skew-normal fits to the transformed copula draws (Section~\ref{sec-copula}).
Under exact calibration, PIT values are uniformly distributed, so their empirical CDF (ECDF) should lie near the diagonal, with systematic deviations indicating miscalibration.

We conduct an SBC check on the benchmark Political Democracy model under two prior settings listed in Table~\ref{tbl-priors}, replicating the analysis of \textcite{merkle2021efficient}.
For each prior setting, SBC replications were run until 500 successful fits were collected.
A replication was discarded if the fit failed, in practice because the Hessian at the mode was not positive definite.
Under the informative priors, all 500 attempts succeeded.
Under the diffuse priors, obtaining 500 successful fits required 540 attempts, giving a failure rate of roughly 7\%.
The check completed in under four minutes.

\protect\phantomsection\label{cell-fig-sbc-ecdf}
\begin{figure}[p]

\centering{

\includegraphics[width=1\linewidth,height=\textheight,keepaspectratio]{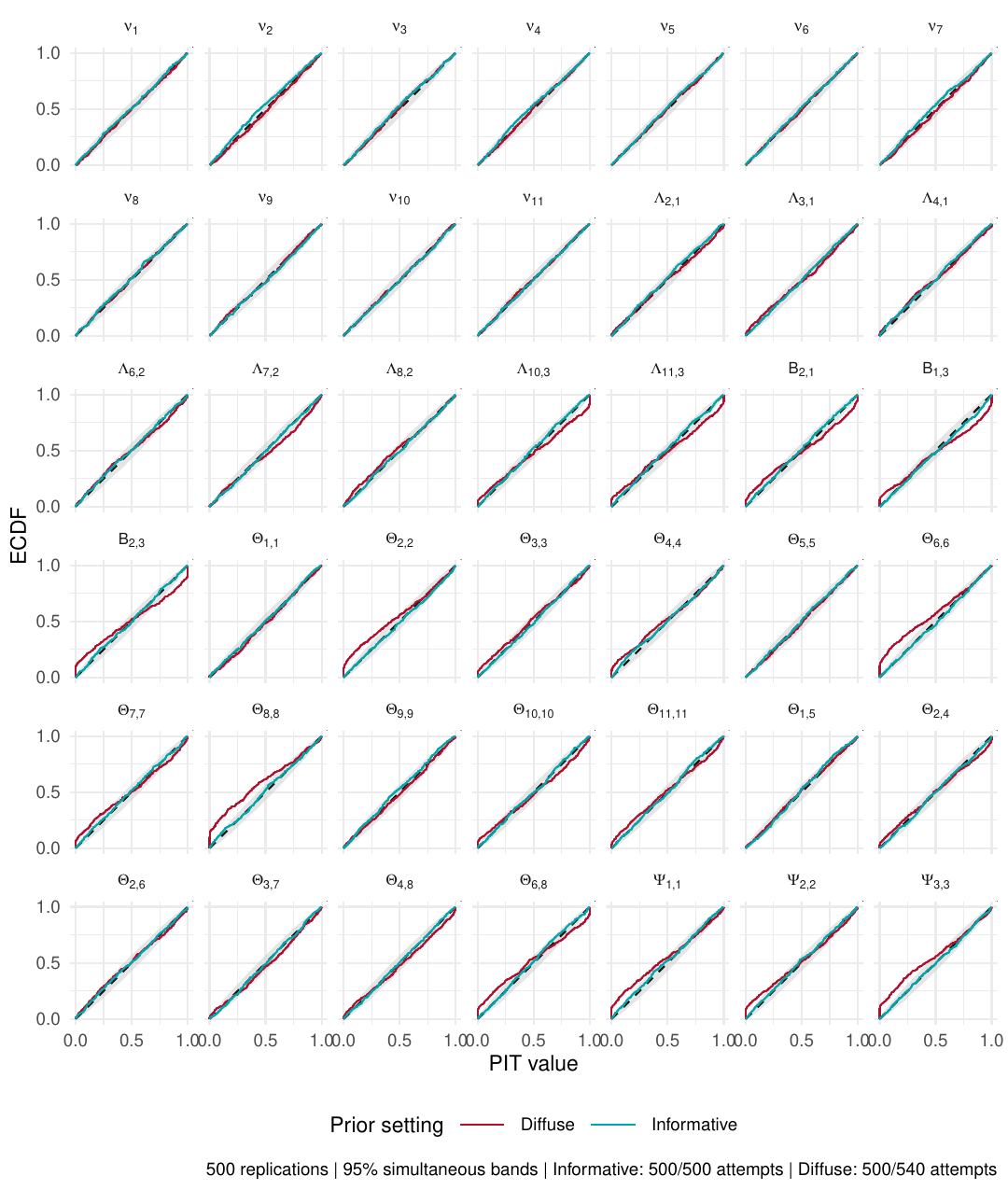}

}

\caption{\label{fig-sbc-ecdf}PIT-ECDF diagnostic plots for all 42 parameters under diffuse and informative priors. The black diagonal line indicates the expected ECDF under perfect calibration, while the shaded grey area represents 95\% simultaneous confidence bands derived from the exact distribution of uniform order statistics. Deviations of the ECDF curves outside the confidence bands indicate miscalibration. The six residual covariances are evaluated on the covariance scale, from the skew-normal fitted to their copula draws.}

\end{figure}%

Figure~\ref{fig-sbc-ecdf} displays PIT-ECDF diagnostic plots for all 42 parameters under both prior settings, overlaid with 95\% simultaneous confidence bands constructed from the exact distribution of uniform order statistics \autocite{sailynoja2022graphical}.
Under the informative priors, the ECDF curves closely follow the uniform reference, indicating excellent calibration.
Under the diffuse priors, departures from the confidence bands occur for some parameters but are generally mild.
This is attributed in part to the SBC protocol, since discarding failed replications alters the joint distribution of generating parameters and data, so the self-consistency identity no longer applies to the retained sample.
The replications with non-positive-definite Hessians are predominantly those with a large generating \(\Psi_{3,3}\), and the \(\eta_3\) block (its connected loadings and coefficients \(B_{1,3}\) and \(B_{2,3}\)) carries the effect.
The remaining departures fall on the variances---most visibly \(\Theta_{6,6}\), \(\Theta_{8,8}\), and the remaining latent variances---which rise above the diagonal near the origin, indicating that the generating value falls in the lower tail of the fitted marginal too often.
For some of these departures, the pattern is consistent with the underestimation of near-zero mass observed in Section~\ref{sec-simulation}.

Beyond this, drawing firmer conclusions is difficult except by comparison with MCMC, where a related phenomenon occurs.
\textcite{merkle2021efficient} report pronounced U-shaped rank histograms across nearly all parameters under the same diffuse priors, which they attribute to the sampler implicitly conditioning on positive-definite covariance configurations within each fit.
In our case, selection operates between replications through the exclusion of failed fits, and the observed departures are mild and concentrated in specific parameters.
Neither pattern alone identifies the inference algorithm as the source of miscalibration.

The broader lesson is one of prior specification.
Diffuse independent priors on SDs and correlations can place substantial mass on implausible or inadmissible covariance configurations, a ``garbage in, garbage out'' problem that afflicts any inference engine.
This underscores the importance of thoughtful prior elicitation in Bayesian
SEM, and motivates the development of principled default priors that concentrate mass on scientifically plausible configurations by construction (see discussion in Section~\ref{sec-discussion}).
Overall, the approximation is well calibrated when the priors are appropriately specified, and its departures under diffuse priors are mild and localised.

\section{Discussion}\label{sec-discussion}

We now situate the proposed framework relative to existing approximate inference strategies, before examining its current limitations and directions for further development.

\subsection{Contrast With Original INLA}\label{contrast-with-original-inla}

The proposed framework shares its conceptual roots with INLA \autocite{rue2009approximate}, yet departs from it in several important respects that are worth clarifying, particularly for readers familiar with that methodology.
As a starting point, both approaches construct a joint Laplace approximation, but INLA targets the conditional latent field whereas our approach targets the parameter posterior.
In the general INLA setting, a potentially high-dimensional latent Gaussian field is retained in the inference scheme, and the crudeness of the joint Gaussian approximation motivates the classical nested Laplace strategy.
The marginal for each hyperparameter is constructed by repeatedly integrating out the latent field at each configuration point, a procedure that can be expensive when the latent field is large.
In our setting, the Gaussian conjugacy of the SEM likelihood permits the latent variables to be integrated out analytically, which removes the repeated integration over the latent field and makes exact (unnormalised) log-posterior evaluations inexpensive.
What remains is a problem of moderate dimension, where the whole of the inference takes place in the \(m\)-dimensional space of \(\boldsymbol\vartheta\).

The VB correction procedure exemplifies this difference in scale.
Within INLA, the VB update targets the latent field, which may number in the thousands, making a low-rank approximation necessary for tractability \autocite{vanniekerk2024lowrank}.
In our framework, the correction operates directly on the \(m\)-dimensional hyperparameter vector, so the full mean shift can be computed without low-rank truncation.

Another consequential difference lies in marginal density fitting.
INLA constructs its parametric marginal approximations by matching low-order derivatives of the approximating family to derivatives of the log-marginal at a small number of evaluation points (typically four or five), a design choice motivated by the high cost of each function evaluation in the nested Laplace scheme.
An alternative within that framework is the asymmetric two-piece Gaussian, which requires only two evaluations \autocite{martins2013bayesian}.
While parsimonious, derivative matching can yield poor fits when the marginal shape departs from what can be captured by a few local derivatives.
By contrast, since exact joint log-posterior evaluations along the scan grid are cheap, we fit a skew-normal curve on a dense grid spanning essentially all of the local Gaussian mass.
This yields substantially better fidelity whenever the marginal is approximately skew-normal.
When it is not, e.g., in the presence of heavy tails or boundary effects, then one may resort to richer constructions such as the nonparametric tail corrections described by \textcite{rue2009approximate}.

Finally, the methodological contributions of this work may be of interest beyond SEM.
The efficient volume correction (Lemma~\ref{lem-eff-vol} and Corollary~\ref{cor-eff-vol}) provides a novel gradient-based scheme for updating Laplace volumes along scan trajectories, with potential applications in INLA and other profiled Laplace approximations.
The volume correction, a device for marginalisation, turns out to be a device for recentring as well.
Proposition~\ref{prp-vb-coherence} identifies the leading mode-to-mean displacement as one quantity recovered jointly by variational optimisation and componentwise by volume-corrected profiling, with agreement to \(O(n^{-2})\).
Modern INLA replaces the nested Laplace stage for the latent field with VB-corrected conditional Gaussians, which can produce skewed marginals through hyperparameter integration \autocite{vanniekerk2023new}.
Our construction, though rooted in the classical nested strategy, thus recovers through marginalisation the leading location correction targeted by the modern variational approach.
The equivalence also provides an internal diagnostic, since the VB-corrected location predicts the fitted skew-normal mean, with discrepancies signalling higher-order effects or numerical error.

\subsection{Contrast With Variational Inference}\label{contrast-with-variational-inference}

Variational inference \autocite[VI;][]{blei2017variational} offers an alternative route to tractable posterior approximation by casting inference as an optimisation problem over a family of candidate densities.
The most common implementation is \emph{mean-field VI}, which assumes posterior independence across parameters, an assumption that is manifestly violated in SEMs, where loadings, regression coefficients, and variance components are typically correlated a posteriori.
\textcite{dang2022fitting} studied VI for confirmatory factor models and found that, while the posterior modes of the marginals were recovered accurately, the posterior SDs were substantially underestimated.
Correcting these required a computationally expensive bootstrap adjustment, which cost much of the speed advantage that motivated VI in the first place.

VI has also been applied to structure selection in factor and multiple-indicators multiple-causes (MIMIC) models, with spike-and-slab priors on unspecified loadings or covariate effects and a variational approximation in place of the Gibbs sampler \autocite{jin2024regularized,jin2025regularized}, although this is a different inferential target from the fixed-structure, full-posterior problem treated here.
For the fixed-structure case, the generic engine is Stan's automatic differentiation VI \autocite[ADVI;][]{kucukelbir2017automatic}, which applies to the models of this paper as is.
Its implementation is still labelled experimental by the Stan developers and, in our experience, sensitive to its convergence settings, but it offers the natural route to a systematic comparison of variational and Laplace-based approximations in SEM.

Our VB approach in Section~\ref{sec-vb} occupies a middle ground.
We use VB-type optimisation, grounded in an information-processing view of Bayesian inference \autocite{zellner1988optimal,knoblauch2022optimizationcentric}, to refine the posterior centre, while relying on Laplace curvature and skew-normal profiling for uncertainty quantification.
The Gaussian copula reconstruction further carries the dependence structure of the joint Hessian into the joint draws, avoiding the independence assumption that limits mean-field VI.

\subsection{Limitations of the Proposed Method}\label{sec-limitations}

The proposed approximation inherits the fundamental character of Laplace-based methods, i.e., accuracy is highest when the posterior is dominated by a single, well-separated interior mode, so that a local quadratic expansion faithfully represents the log-posterior in the region containing most of the posterior probability.
The primary failure mode is a loss of positive definiteness of the Hessian \(\mathbf H\) at the candidate mode, which signals that the local curvature is insufficient to define a proper Gaussian approximation.
Such failure can reflect weakly identified directions along which distinct parameter combinations receive similar posterior support.
These identification difficulties are not unique to our framework, as they also affect maximum likelihood estimation and can hinder MCMC sampling \autocite{lee2004evaluation}.

Failures of this kind often point to identifiability or data adequacy issues, so the usual remedies are to increase the sample size or strengthen the priors \autocite{ulitzsch2023alleviating}.
In our studies, estimation error decreased with increasing sample size (Section~\ref{sec-simulation}), while the SBC checks (Section~\ref{sec-sbc}) revealed that roughly 7\% of replications under diffuse priors failed to produce a positive-definite Hessian, whereas none failed under informative priors.

Even when \(\mathbf H\) is positive definite, proximity to a variance boundary can compromise the accuracy of the fitted marginals.
With variances entering on the log scale (Table~\ref{tbl-priors}), their fitted marginals are log-skew-normal on the natural scale, ensuring positivity and accommodating long right tails and mode-to-mean gaps.
This form can nevertheless understate posterior mass near zero despite accurate central summaries, as Section~\ref{sec-simulation} demonstrates.
Inference sensitive to these lower-tail probabilities should therefore be checked against MCMC.

Another limitation is that increasing \(m\) may make both mode-finding and marginal profiling more demanding.
Dense Hessian operations become more expensive as \(m\) grows, and each additional parameter requires a marginal scan.
Although scans can run in parallel and the conjugate Gaussian structure permits posterior evaluations without \(O(n)\) work, larger parameter spaces may still complicate mode-finding and curvature estimation.
In typical applied psychometric settings involving three to five latent factors, \(m\) ranges from roughly 20 to 60; more complex specifications with additional cross-loadings, correlated residuals, or multigroup analysis may push \(m\) to 100--200.
Even for the more complex models examined in Section~\ref{sec-simulation} (Scenario S5), these costs remained manageable, while MCMC was substantially slower.
Much larger models may nevertheless require alternative strategies, such as low-rank or block-structured approximations to the \(m \times m\) Hessian.

\subsection{Further Improvements}\label{further-improvements}

Several extensions of the present framework merit investigation.
First, the linear volume correction of Section~\ref{sec-efficient-volume} could be extended to second order by including a quadratic term \(\tfrac{1}{2}\gamma_j''\,t^2\) in the Taylor expansion of \(\gamma_j(t)\).
This would accommodate nonlinear variation in the conditional spread along the scan direction, which may improve marginal accuracy for parameters whose nuisance curvature changes appreciably away from the mode.
Computationally, the second derivative of the log-determinant involves two traces, \(\operatorname{tr}\big\{\mathbf J''(0)\big\}\) and \(\operatorname{tr}\big\{\mathbf J'(0)^2\big\}\), with an additional Schur complement correction analogous to the first-order case.
While the first term admits the same gradient-based evaluation used for the first-order correction, the second term involves a Frobenius norm of a third-derivative contraction that resists a comparably cheap deterministic reduction.
Whether the resulting accuracy gains justify the added complexity remains an open question.

Relatedly, automatic differentiation through a compiled backend could replace the finite-difference calculations of the Hessian and volume slopes.
Since the gradient is already available in closed form, this would remove step-size dependence from the higher derivatives, with potential gains in numerical robustness and computational efficiency as model dimension grows.

Second, the variational correction currently updates only the posterior location while holding the covariance fixed at the Laplace value.
A natural extension is to optimise both the mean and the covariance simultaneously, as proposed by \textcite{dutta2026scalable}, which could improve calibration when the Laplace covariance is a poor approximation to the posterior spread.

Third, the prior specification adopted here follows standard default choices in Bayesian SEM software.
More principled alternatives, such as penalised complexity priors \autocite{simpson2017penalising}, particularly for covariance structures \autocite{freni2025graphical,freni2026informative}, could improve both interpretability and computational stability.
Such priors would also avoid the non-positive-definite inconsistencies that can arise from specifying independent priors on SDs and correlations.
Heavier-tailed prior families (e.g., Student-\(t\)) would additionally guard against conflict between prior and data, since the posterior can then follow the likelihood where the two disagree.

Finally, the most substantive direction for future work is extending the framework beyond Gaussian likelihoods.
The analytic marginalisation of latent variables that underpins our approach relies on conjugacy, which is generally lost with binary, ordinal, count, or bounded continuous responses.
Following the original INLA construction in (\ref{eq-sem-inla}), a Laplace approximation could yet again integrate out the latent variables at each parameter evaluation, supplying an approximate marginal likelihood for the subsequent joint and marginal constructions.
The resulting approximate marginal likelihood would then support the existing procedures for mode-finding, skew-normal profiling, variational correction, and copula construction.
This extension would therefore add one approximation layer while retaining the present architecture, with the success of INLA providing a strong precedent for its feasibility.

\section{Conclusion}\label{sec-conclusion}

This article presented a framework for fast, deterministic Bayesian inference in linear Gaussian SEMs, built on a sequence of Laplace-type approximations adapted to the SEM setting.
By analytically integrating out latent variables and operating entirely in the marginal likelihood space of the model parameters, the approach avoids the cost of high-dimensional latent-field approximations and delivers accurate posterior marginals in a fraction of the time required by MCMC.
On the Political Democracy \autocite{bollen1989structural} benchmark, the full approximation completed in 1.5 seconds, achieving JS similarity of 98.5--100\% against MCMC across all 42 parameters, with similarly close agreement for factor scores and model-fit summaries.
The marginal ablation study showed how the profiling components work together to improve agreement with the MCMC reference.
Across the simulation and calibration studies, accuracy improved with sample size and remained high as model dimension increased, with difficulties concentrated in weakly identified parameters, near-zero variance tails, and configurations arising under diffuse priors.
The studies also showed that the variational shift and marginal fitting residuals can provide useful warning signs of deteriorating approximation accuracy.

We see this work not as a replacement for MCMC, but as a complement to it within the Bayesian SEM workflow.
The speed of the proposed approximation makes it well suited to the iterative cycle of model specification, evaluation, and revision that characterises applied psychometric practice.
Researchers can rapidly compare competing measurement or structural specifications, assess sensitivity to prior choices, and identify promising candidates before committing to a full MCMC run for final inference.
Once a model has been selected, MCMC remains available as a gold-standard reference.
The operationalisation of these workflows within the proposed methodology will be described in an accompanying applications- and software-oriented paper.

\section*{Acknowledgements}\label{acknowledgements}
\addcontentsline{toc}{section}{Acknowledgements}

We thank the four anonymous reviewers, whose comments substantially strengthened this work.
This publication is based upon work supported by the King Abdullah University of Science and Technology (KAUST) Research Funding Office under Award No.~URF/1/6921-01-01.

\section*{Data Availability}\label{data-availability}
\addcontentsline{toc}{section}{Data Availability}

The empirical dataset used in this article (Political Democracy) is publicly accessible within the R lavaan package \autocite{rosseel2012lavaan}.
All code required to fully reproduce the results is deposited in an open repository at \url{https://osf.io/arqmh/}.
Supplementary Material, which reports numerical checks and the diagnostic calibration for the simulation study, is available in the same repository.

\section*{References}\label{references}
\addcontentsline{toc}{section}{References}

\printbibliography[heading=none]

\newpage{}

\appendix

\section{\texorpdfstring{Proof of Lemma~\ref{lem-eff-vol} (Volume-Slope Decomposition)}{Proof of Lemma~ (Volume-Slope Decomposition)}}\label{proof-of-lem-eff-vol-volume-slope-decomposition}

Since \(\boldsymbol\vartheta(0) = \boldsymbol\vartheta^*\) by construction,
\[
\mathbf J(0) = \mathbf L^\top \mathbf H(\boldsymbol\vartheta^*)\,\mathbf L = \mathbf L^\top (\mathbf L\mathbf L^\top)^{-1}\mathbf L = \mathbf L^\top \mathbf L^{-\top}\mathbf L^{-1}\mathbf L = \mathbf I_m.
\]
Further, as \(\mathbf v_j = \boldsymbol\Omega_{\cdot j}/\sqrt{\Omega_{jj}}\) and \(\boldsymbol\Omega = \mathbf L\mathbf L^\top\), we have \(\mathbf w_j = \mathbf L^{-1}\mathbf v_j\) and, using \(\boldsymbol\Omega^{-1}\boldsymbol\Omega_{\cdot j} = \mathbf e_j\) (the \(j\)th canonical basis vector),
\[
\|\mathbf w_j\|^2 = \mathbf v_j^\top \boldsymbol\Omega^{-1}\mathbf v_j = \boldsymbol\Omega_{\cdot j}^\top \mathbf e_j/\Omega_{jj} = \Omega_{jj}/\Omega_{jj} = 1.
\]

Partition the matrix \(\mathbf H\) as
\[
\mathbf H = \begin{pmatrix}
H_{jj} & \mathbf h_j^\top \\
\mathbf h_j & \mathbf H_{-j}
\end{pmatrix},
\]
and let \(S_j = H_{jj} - \mathbf h_j^\top \mathbf H_{-j}^{-1}\mathbf h_j\) denote the Schur complement of \(\mathbf H_{-j}\) in \(\mathbf H\).
Using the Schur complement identity \(|\mathbf H_{-j}| = |\mathbf H|/S_j\), we decompose
\[
\gamma_j(t) = -\frac12\log|\mathbf H(t)| + \frac12\log S_j(t).
\]

For the first term, notice that \(-\frac12\log|\mathbf H(t)| = -\frac12\log|\mathbf J(t)| + \log|\mathbf L|\).
Differentiating with respect to \(t\), the constant term \(\log|\mathbf L|\) vanishes, and thus we may evaluate the derivative in the whitened frame.
At \(t=0\), Jacobi's formula \autocite[Chapter 8, Theorem 2]{magnus1999matrix} gives
\[
-\frac12 \frac{d}{dt}\log|\mathbf J(t)|\bigg|_{t=0} = -\frac12 \operatorname{tr}\left\{\left.\frac{d\mathbf J(t)}{dt}\right|_{t=0}\right\}.
\]

For the second term, differentiating the Schur complement identity \((\mathbf H^{-1})_{jj} = 1/S_j\) using \(d \mathbf H^{-1} = -\mathbf H^{-1}(d\mathbf H)\mathbf H^{-1}\) \autocite[Chapter 8, Theorem 3]{magnus1999matrix}, and recalling that \(\mathbf H^{-1} = \boldsymbol\Omega\) we obtain
\[
\frac{d}{dt}(\mathbf H^{-1})_{jj}\bigg|_{t=0} = -\boldsymbol\Omega_{\cdot j}^\top \mathbf H'(0)\,\boldsymbol\Omega_{\cdot j},
\]
where \(\mathbf H'(0) = \frac{d\mathbf H(t)}{dt}\big|_{t=0}\).
Since \(\log S_j = -\log(\mathbf H^{-1})_{jj}\), the chain rule gives \(\frac{d}{dt}\log S_j = \frac{-(d/dt)(\mathbf H^{-1})_{jj}}{(\mathbf H^{-1})_{jj}}\), which at \(t=0\) yields
\[
\frac{d}{dt}\log S_j(t)\bigg|_{t=0} = \frac{\boldsymbol\Omega_{\cdot j}^\top \mathbf H'(0)\,\boldsymbol\Omega_{\cdot j}}{\Omega_{jj}} = \mathbf v_j^\top \mathbf H'(0)\,\mathbf v_j = \mathbf w_j^\top \mathbf J'(0)\,\mathbf w_j,
\]
using the fact that \(\mathbf v_j = \boldsymbol\Omega_{\cdot j}/\sqrt{\Omega_{jj}}\) and \(\mathbf w_j = \mathbf L^{-1}\mathbf v_j\).
Combining both terms yields the result.

\section{\texorpdfstring{Proof of Corollary~\ref{cor-eff-vol} (Gradient Representation of the Volume Slope)}{Proof of Corollary~ (Gradient Representation of the Volume Slope)}}\label{proof-of-cor-eff-vol-gradient-representation-of-the-volume-slope}

For any direction \(\mathbf u\) not depending on \(t\), the identity \(\frac{d}{d\mathbf u}\mathbf g = \mathbf H\mathbf u\) gives \(\mathbf u^\top\mathbf H\,\mathbf u = \mathbf u^\top\frac{d}{d\mathbf u}\mathbf g\), and since \(\mathbf u\) is a fixed prefactor the \(t\)-derivative passes through to give
\[
\mathbf u^\top\mathbf H'(0)\,\mathbf u = \frac{d}{dt}\!\left[\mathbf u^\top\frac{d}{d\mathbf u}\mathbf g\right]_{t=0}.
\]
The trace in (\ref{eq-vol-slope}) expands as \(\operatorname{tr}\{\mathbf J'(0)\} = \sum_{k=1}^m \mathbf L_{\cdot k}^\top\mathbf H'(0)\,\mathbf L_{\cdot k}\), and the Schur term satisfies \(\mathbf w_j^\top\mathbf J'(0)\,\mathbf w_j = \mathbf v_j^\top\mathbf H'(0)\,\mathbf v_j\) using \(\mathbf w_j = \mathbf L^{-1}\mathbf v_j\).
Applying the identity to each of the \(m+1\) directions \(\mathbf u \in \{\mathbf L_{\cdot 1},\dots,\mathbf L_{\cdot m},\mathbf v_j\}\) and substituting into (\ref{eq-vol-slope}) yields (\ref{eq-vol-slope-gradient}).

\section{\texorpdfstring{Proof of Proposition~\ref{prp-vb-coherence} (Equivalence of Location Corrections)}{Proof of Proposition~ (Equivalence of Location Corrections)}}\label{proof-of-prp-vb-coherence-equivalence-of-location-corrections}

Write \(\ell(\boldsymbol\vartheta)=\log\pi(\boldsymbol\vartheta\mid\mathbf y)\) for the log-posterior.
The regularity conditions of \textcite{tierney1986accurate} stated in Section~\ref{sec-preliminaries} are assumed, so that the derivatives of \(\ell\) near its interior mode \(\boldsymbol\vartheta^*\) are \(O(n)\) and \(\boldsymbol\Omega=\mathbf H^{-1}=O(n^{-1})\).
It is also assumed that the working posterior is positive on \(\mathbb R^m\).

Fix \(j\in\{1,\dots,m\}\), and put
\[
\sigma^2=\Omega_{jj},
\qquad
\mathbf v=\mathbf v_j=\frac{\boldsymbol\Omega_{\cdot j}}{\sigma},
\qquad
\mathbf A=\left.\frac{d}{dt}\nabla^2\ell(\boldsymbol\vartheta^*+t\mathbf v)\right|_{t=0} = - \mathbf H'(0).
\]
Thus \(\mathbf A\) describes the change in log-posterior curvature along the scan \(\vartheta_j(t)=\vartheta_j^*+\sigma t\).
Pertinent to this proof is the distribution of \(\boldsymbol\vartheta - \boldsymbol\vartheta^* =: \boldsymbol\epsilon \sim\N_m(\mathbf 0,\boldsymbol\Omega)\), whose conditional covariance given \(\epsilon_j\) is \(\mathbf C=\boldsymbol\Omega-\mathbf v\mathbf v^\top\).

Since \(\mathbf v^\top\mathbf H\mathbf v=1\), the local expansion of the corrected log profile (\ref{eq-corrected-ordinates}) is
\[
h_j(t)
=
c-\tfrac12t^2+
\overbrace{\gamma_j'\,t+\tfrac16 s\,t^3
+r(t)+O(n^{-3/2})}^{\Delta(t)},
\qquad s=\mathbf v^\top\mathbf A\mathbf v,
\]
where \(\gamma_j'\) and \(s\) are \(O(n^{-1/2})\), and \(r\) is an even quartic term of order \(n^{-1}\).
Let \(p_j\propto\exp(h_j)\), and consider its mean
\[
\operatorname{E}_{p_j}(t)
=
\frac{\int t\,e^{-t^2/2}e^{\Delta(t)}\,dt}{\int e^{-t^2/2}e^{\Delta(t)}\,dt}
=
\frac{\operatorname{E}\bigl\{Z e^{\Delta(Z)}\bigr\}}{\operatorname{E}\bigl\{e^{\Delta(Z)}\bigr\}},
\qquad Z\sim\N(0,1),
\]
obtained by factoring out the Gaussian kernel.
Expanding \(e^\Delta=1+\Delta+\tfrac12\Delta^2+\cdots\), and using \(\operatorname{E}(Z^2)=1\) and \(\operatorname{E}(Z^4)=3\), gives
\[
\operatorname{E}_{p_j}(t)
=
\frac{
\gamma_j'\operatorname{E}(Z^2)+\tfrac16 s\operatorname{E}(Z^4)+O(n^{-3/2})
}{
1+O(n^{-1})
}
=
\gamma_j'+\tfrac12 s+O(n^{-3/2}).
\]
The \(O(n^{-1/2})\) terms in the denominator vanish because they are odd functions of \(Z\), and similarly for the even \(O(n^{-1})\) terms in the numerator after multiplication by \(Z\).
Since the numerator is itself \(O(n^{-1/2})\), dividing by \(1+O(n^{-1})\) perturbs it only at \(O(n^{-3/2})\).
Gaussian integrability of the polynomial terms in \(\Delta\), together with the stated Laplace tail condition, permits averaging the expansion term by term.

Returning to the original scale, we therefore obtain
\begin{equation}\protect\phantomsection\label{eq-marginal-shift}{
\begin{aligned}[b]
\bar\vartheta_j-\vartheta_j^*
=\sigma\left\{\gamma_j'+\frac12s+O(n^{-3/2})\right\} 
&=
\frac{\sigma}{2}
\left\{\operatorname{tr}(\mathbf A\mathbf C)+\mathbf v^\top\mathbf A\mathbf v\right\}
+O(n^{-2})\\
&=
\frac{\sigma}{2}\operatorname{tr}(\mathbf A\boldsymbol\Omega)
+O(n^{-2}),
\end{aligned}
}\end{equation}
where, by Lemma~\ref{lem-eff-vol}, the volume slope in original coordinates is \(\gamma_j'=\tfrac12\operatorname{tr}(\mathbf A\mathbf C)\).
Now consider the VB objective (\ref{eq-vb-objective}), which under \(\boldsymbol\vartheta=\boldsymbol\vartheta^*+\boldsymbol\delta+\boldsymbol\epsilon\) reads
\[
\mathcal L(\boldsymbol\delta)=\operatorname{E}_{\boldsymbol\epsilon}\bigl\{\ell(\boldsymbol\vartheta^*+\boldsymbol\delta+\boldsymbol\epsilon)\bigr\}.
\]
Assuming \(\hat{\boldsymbol\delta}=O(n^{-1})\), expanding \(\nabla\ell\) about \(\boldsymbol\vartheta^*\) under the expectation, and projecting the stationarity condition onto \(\mathbf v\) gives
\[
\begin{aligned}
0
&=
\mathbf v^\top\operatorname{E}_{\boldsymbol\epsilon}\bigl\{\nabla\ell(\boldsymbol\vartheta^*+\hat{\boldsymbol\delta}+\boldsymbol\epsilon)\bigr\}\\
&=
\mathbf v^\top\operatorname{E}_{\boldsymbol\epsilon}\left\{
-\mathbf H(\hat{\boldsymbol\delta}+\boldsymbol\epsilon)
+\tfrac12\nabla^3\ell(\boldsymbol\vartheta^*)[\hat{\boldsymbol\delta}+\boldsymbol\epsilon,\hat{\boldsymbol\delta}+\boldsymbol\epsilon]
\right\}+O(n^{-3/2})\\
&=
-\frac{\hat\delta_j}{\sigma}
+\tfrac12\operatorname{E}(\boldsymbol\epsilon^\top\mathbf A\boldsymbol\epsilon)
+O(n^{-3/2}).
\end{aligned}
\]
Here \(\nabla^3\ell(\boldsymbol\vartheta^*)[\mathbf u,\mathbf u]\) is the vector with \(a\)th component \(\sum_{b,c}\partial_{abc}\ell(\boldsymbol\vartheta^*)u_bu_c\), and multiplying by \(\mathbf v^\top\) gives \(\mathbf u^\top\mathbf A\mathbf u\).
We also use \(\mathbf H\mathbf v=\mathbf e_j/\sigma\) and the vanishing of odd centred Gaussian moments.
The quadratic term in \(\hat{\boldsymbol\delta}\) and the expected higher terms are \(O(n^{-3/2})\).
Since \(\operatorname{E}(\boldsymbol\epsilon^\top\mathbf A\boldsymbol\epsilon)=\operatorname{tr}(\mathbf A\boldsymbol\Omega)\),
\[
\hat\delta_j
=
\frac{\sigma}{2}\operatorname{tr}(\mathbf A\boldsymbol\Omega)+O(n^{-2}).
\]
Comparison with (\ref{eq-marginal-shift}) proves
\[
\bar\vartheta_j
=
\vartheta_j^*+\hat\delta_j+O(n^{-2}).
\]

It remains to prove that \(\hat{\boldsymbol\delta}=O(n^{-1})\).
Consider \(q_{\boldsymbol\delta}=\N_m(\boldsymbol\vartheta^*+\boldsymbol\delta,\boldsymbol\Omega)\).
The Taylor expansion of \(\ell\) and the exact log-density of the Laplace Gaussian \(q_{\mathbf0}\) give, respectively,
\[
\begin{alignedat}{2}
\ell(\boldsymbol\vartheta^*+\boldsymbol\epsilon)
&=\ell(\boldsymbol\vartheta^*)
&&{}-\tfrac12\boldsymbol\epsilon^\top\boldsymbol\Omega^{-1}\boldsymbol\epsilon
+\tfrac16\nabla^3\ell(\boldsymbol\vartheta^*)[\boldsymbol\epsilon,\boldsymbol\epsilon,\boldsymbol\epsilon]
+O(n\|\boldsymbol\epsilon\|^4),\\
\log q_{\mathbf0}(\boldsymbol\vartheta^*+\boldsymbol\epsilon)
&=\log q_{\mathbf0}(\boldsymbol\vartheta^*)
&&{}-\tfrac12\boldsymbol\epsilon^\top\boldsymbol\Omega^{-1}\boldsymbol\epsilon.
\end{alignedat}
\]
Subtracting the first from the second cancels the quadratic terms, while the \(O(n^{-1})\) relative error of Laplace's method gives \(\log q_{\mathbf 0}(\boldsymbol\vartheta^*)-\ell(\boldsymbol\vartheta^*)=O(n^{-1})\).
Hence, locally,
\[
\log\frac{q_{\mathbf0}(\boldsymbol\vartheta^*+\boldsymbol\epsilon)}{\pi(\boldsymbol\vartheta^*+\boldsymbol\epsilon\mid\mathbf y)}
=-\tfrac16\nabla^3\ell(\boldsymbol\vartheta^*)[\boldsymbol\epsilon,\boldsymbol\epsilon,\boldsymbol\epsilon]
+O\!\left(n\|\boldsymbol\epsilon\|^4+n^{-1}\right).
\]
The cubic term has zero Gaussian expectation, while \(\operatorname{E}\|\boldsymbol\epsilon\|^4=O(n^{-2})\).
Again, Gaussian integrability and the Laplace tail control permit this.
Therefore,
\[
D_{\mathrm{KL}}(q_{\mathbf0} \, \Vert \, \pi)
=O\left(n\operatorname{E}\|\boldsymbol\epsilon\|^4+n^{-1}\right)
=O(n^{-1}).
\]
By variational optimality,
\[
D_{\mathrm{KL}}(q_{\hat{\boldsymbol\delta}} \, \Vert \, \pi)\le D_{\mathrm{KL}}(q_{\mathbf0} \, \Vert \, \pi)=O(n^{-1}).
\]
Writing \(H\) for the Hellinger distance, the triangle inequality and \(H^2(p,q)\le\tfrac12D_{\mathrm{KL}}(p \, \Vert \, q)\) give
\[
\begin{aligned}
H^2(q_{\mathbf0},q_{\hat{\boldsymbol\delta}})
&\le \bigl\{H(q_{\mathbf0},\pi)+H(q_{\hat{\boldsymbol\delta}},\pi)\bigr\}^2\\
&\le 2H^2(q_{\mathbf0},\pi)+2H^2(q_{\hat{\boldsymbol\delta}},\pi)\\
&\le D_{\mathrm{KL}}(q_{\mathbf0} \, \Vert \, \pi)+D_{\mathrm{KL}}(q_{\hat{\boldsymbol\delta}} \, \Vert \, \pi)
=O(n^{-1}).
\end{aligned}
\]
For these two Gaussians with common covariance \(\boldsymbol\Omega\),
\[
H^2(q_{\mathbf0},q_{\hat{\boldsymbol\delta}})
=
1-\exp \left(-\tfrac18\hat{\boldsymbol\delta}^{\top}\boldsymbol\Omega^{-1}\hat{\boldsymbol\delta}\right)
=O(n^{-1}).
\]
Thus \(\hat{\boldsymbol\delta}^{\top}\boldsymbol\Omega^{-1}\hat{\boldsymbol\delta}=O(n^{-1})\), which gives \(\|\hat{\boldsymbol\delta}\|^2=O(n^{-2})\), and hence as required, \(\hat{\boldsymbol\delta}=O(n^{-1})\).

Three remarks are in order.
First, the quantity \(\tfrac{\sigma}{2}\operatorname{tr}(\mathbf A\boldsymbol\Omega)\) is the leading \(O(n^{-1})\) term \(b_j\) in the mode-to-mean expansion of \textcite[Eq. 6]{lindley1980approximate}, for which \(\operatorname{E}(\vartheta_j\mid\mathbf y)=\vartheta_j^*+b_j+O(n^{-2})\).
VB optimisation thus recovers Lindley's first-order correction, giving the variational shift a classical asymptotic justification.
The profile approximation independently recovers the same term.

Second, the expansion of \(h_j\) is that of the simplified Laplace approximation of \textcite[Section 3.2.3]{rue2009approximate}, with \(\gamma_j'\) and \(s\) as their \(\gamma^{(1)}_j\) and \(\gamma^{(3)}_j\).
Their skew-normal is fitted to have mean \(\gamma^{(1)}_j\) (along with unit variance and third derivative \(\gamma^{(3)}_j\) at the mode), whereas the mean of \(p_j\propto\exp(h_j)\) on the standardised scan scale is \(\gamma_j'+\tfrac12 s\) to the order shown.
The omitted term contributes \(\sigma s/2=O(n^{-1})\) on the original scale, so their skew-normal mean misses \(b_j\) by that amount.
We instead fit the corrected ordinates (\ref{eq-corrected-ordinates}), which leaves the skew-normal mean free to reflect both contributions.

Third and last, the proposition concerns the continuous profile density in \(t\) and the exact VB expectation.
The least-squares fit on a finite grid, finite-difference and curve-fitting errors, and the RQMC evaluation of the VB objective are separate numerical errors; no order in \(n\) is claimed for them, and they also limit agreement in practice.

\section{Volume Correction by Parameter Class}\label{volume-correction-by-parameter-class}

\setcounter{table}{0}
\renewcommand{\thetable}{\Alph{section}\arabic{table}}

Table~\ref{tbl-gamma1-appendix} accompanies the discussion in Section~\ref{sec-ablation}, reporting the effect of the volume correction disaggregated by parameter class.

\begin{table}[h]

\caption{\label{tbl-gamma1-appendix}Effect of the volume correction \(\gamma_j'\) on the fitted marginals, by parameter class. The three fits are construction (iii) of Table~\ref{tbl-ablation-ladder} and differ only in how the correction is obtained: (a) omitted entirely, (b) by the gradient-only scheme of Corollary~\ref{cor-eff-vol}, the default, and (c) from a full numerical Hessian. \(N\) denotes the number of parameters in each class.}

\centering{

\fontsize{10.0pt}{12.0pt}\selectfont
\begin{tabular*}{\linewidth}{@{\extracolsep{\fill}}l|rrrrrrr}
\toprule
 &  & \multicolumn{3}{c}{{JS discrepancy (\%)}} & \multicolumn{3}{c}{{\(\Delta_{\mathrm{Mean},j}\)}} \\ 
\cmidrule(lr){3-5} \cmidrule(lr){6-8}
Parameter & \(N\) & (a) & (b) & (c) & (a) & (b) & (c) \\ 
\midrule\addlinespace[2.5pt]
\multicolumn{1}{l}{{Observed intercepts}} & 11 & 0.05 & 0.05 & 0.05 & 0.011 & 0.011 & 0.011 \\ 
\multicolumn{1}{l}{{Loadings}} & 8 & 0.84 & 0.24 & 0.24 & 0.177 & 0.053 & 0.052 \\ 
\multicolumn{1}{l}{{Regressions}} & 3 & 0.48 & 0.17 & 0.17 & 0.118 & 0.023 & 0.023 \\ 
\multicolumn{1}{l}{{Residual variances}} & 11 & 2.31 & 0.24 & 0.24 & 0.308 & 0.044 & 0.044 \\ 
\multicolumn{1}{l}{{Residual correlations}} & 6 & 0.28 & 0.10 & 0.10 & 0.101 & 0.022 & 0.023 \\ 
\multicolumn{1}{l}{{Latent variances}} & 3 & 0.82 & 0.56 & 0.56 & 0.129 & 0.082 & 0.082 \\ 
\multicolumn{1}{l}{{\bfseries Overall}} & {\bfseries 42} & {\bfseries 0.91} & {\bfseries 0.19} & {\bfseries 0.19} & {\bfseries 0.149} & {\bfseries 0.035} & {\bfseries 0.035} \\ 
\bottomrule
\end{tabular*}
\begin{minipage}{\linewidth}
\vspace{.05em}
Total wall-clock time 0.43, 1.12 and 1.87 s respectively.\\
\end{minipage}

}

\end{table}%

\end{document}